\documentclass[galaxies,article,accept,pdftex,moreauthors]{Definitions/mdpi}
\usepackage{comment}
\firstpage{1}
\pubvolume{1}
\issuenum{1}
\articlenumber{0}
\pubyear{2025}
\copyrightyear{2025}
\externaleditor{ }
\datereceived{ }
\daterevised{ } 
\dateaccepted{ }
\datepublished{ }

\hreflink{https://doi.org/} 

\Title{Unveiling the Variability and Chemical Composition of AL\,Col}
\TitleCitation{ {Unveiling} 
 the Variability and Chemical Composition of AL\,Col}

\newcommand{\tess}{{\it TESS}}
\newcommand{\kepler}{{\it Kepler}}
\newcommand{\brite}{{\it BRITE}}

\newcommand{\teff}{$T_\textrm{eff}$}
\newcommand{\logg}{$\log g$}

\newcommand{\kms}{km\,s$^{-1}$}
\newcommand{\vsini}{$\upsilon\sin i$}

\def\spose#1{\hbox to 0pt{#1\hss}}

\Author{
 {Surath C. Ghosh} 
     $^{1,2,}$*\orcidA{},
Santosh Joshi      $^{1}$\orcidB{},
Samrat Ghosh       $^{1}$\orcidC{},
Athul Dileep       $^{1}$\orcidD{},
Otto Trust         $^{3,4}$\orcidG{},
Mrinmoy Sarkar     $^{1}$\orcidE{},
\linebreak   {Jaime Andrés Rosales Guzmán} 
      $^{5}$\orcidF{},
 {Nicolás Esteban Castro-Toledo} 
          $^{5}$\orcidH{},
Oleg Malkov        $^{6}$\orcidI{},
Harinder P. Singh  $^{7}$\orcidJ{},
Kefeng Tan         $^{8}$\orcidK{} and
Sarabjeet S. Bedi  $^{2}$
}

\AuthorNames{
Surath C. Ghosh,   
Santosh Joshi,   
Samrat Ghosh,
Athul Dileep, 
Otto Trust,
Mrinmoy Sarkar,  
J. A. Rosales, 
N. Castro,
Oleg Malkov, 
Harinder P. Singh, 
Kefeng Tan and        
Sarabjeet S. Bedi 
}

\AuthorCitation{
Ghosh, S.C.;  
Joshi, S.; 
Ghosh, S.; 
Dileep, A.;  
Trust, O.; 
Sarkar, M.;   
Rosales, J.A.;
Castro, N.;   
Malkov, O.; 
Singh, H.P.;
et al.}
    
\address{%
$^{1}$ \quad Aryabhatta Research Institute of Observational Sciences (ARIES),  {Nainital, 263001,} 
 India;  {santosh@aries.res.in (S.J.); samrat.ghosh@aries.res.in (S.G.); dileep@aries.res.in (A.D.); mrinmoysarkar@aries.res.in (M.S.)} 
\\
$^{2}$ \quad  {Department Of Computer Science And Information Technology, MJP Rohilkhand University,} 
  {Bareilly, 243006,} 
 India; ssbedi@mjpru.ac.in \\
$^{3}$ \quad Department of Physics, Mbarara University of Science and Technology,  {Mbarara, 76,} 
 Uganda;  {otrust@must.ac.ug} 
\\
$^{4}$ \quad Department of Physics, Faculty of Science, Kyambogo University,  {Kyambogo, 1,} 
 Uganda\\
$^{5}$ \quad  {Departamento de Física, Universidad de Concepción (UdeC),} 
  {Concepcion, 4070412,} 
 Chile; jrosales@astro-udec.cl (J.A.R.); \linebreak   {ncastro2019@udec.cl (N.C.)} 
\\
$^{6}$ \quad Institute of Astronomy of the Russian Academy of Sciences (INASAN),  {Moscow, 119017,} 
 Russia; malkov@inasan.ru\\
$^{7}$ \quad  {Department of Physics and Astrophysics, University of Delhi,} 
  {India; 110007,} 
 hpsingh.du@gmail.com\\
$^{8}$ \quad National Astronomical Observatories, Chinese Academy of Sciences (NAOC),  {Beijing, 100101,} 
 China; tan@nao.cas.cn\\
}  

\corres{Correspondence:  {su5virat@gmail.com or surath@aries.res.in} 
}

\abstract{In this study, we present analysis of  {{\bf\tess\ }} 
 photometry, spectral energy distribution (SED), high-resolution spectroscopy, and spot modeling of the $\alpha^2$ CVn-type star AL\,Col (HD\,46462). The primary objective is to determine its fundamental physical parameters and  investigate its surface activity characteristics.
Using \tess\ short-cadence (120\,s) SAP flux, we identified a rotational frequency of 0.09655\,$\mathrm{d}^{-1}$ ($P_\mathrm{rot}=10.35733$\,d). Wavelet analysis reveals that while the amplitudes of the harmonic components vary over time, the strength of the primary rotational frequency remains stable. A SED analysis of multi-band photometric data yields an effective temperature ($T_\mathrm{eff}$) of  {11,750\,K.} 
 High-resolution spectroscopic observations covering wavelengthrange 4500--7000\,\AA\  provide refined estimates of \teff\, =\, 13,814\, $\pm$\, 400\,K, \logg\,=\, 4.09\, $\pm$\, 0.08\,dex, and \vsini\, =\, 16 $\pm$ 1\,\kms. Abundance analysis shows solar-like composition of O\,\textsc{ii}, Mg\,\textsc{ii}, S\,\textsc{ii}, and Ca\,\textsc{ii}, while helium is under-abundant by 0.62\,dex. Rare earth elements (REEs) exhibit over-abundances of up to 5.2\,dex, classifying the star as an Ap/Bp-type star. AL\,Col has a radius of $R = 3.74\,\pm\,0.48{\rm R_{\odot}}$, with its H--R diagram position estimating a mass of $M = 4.2\,\pm\,0.2{\rm M_{\odot}}$ and an age of $0.12\,\pm\,0.01$ Gyr, indicating that the star has slightly evolved from the main sequence. The \tess\ light curves were modeled using a three-evolving-spot configuration, suggesting the presence of differential rotation. This star is a promising candidate for future investigations of magnetic field diagnostics and the vertical stratification of chemical elements in its atmosphere.
}

\keyword{  {stars---chemically peculiar;} 
 He weak {;}
 techniques---TESS photometry; high resolution spectroscopy; abundance {;} individual---AL\,Col {;} variable---$\alpha$$^2$CVn}

\begin{document}




\section{Introduction}
\label{Sec: Sec. 1}

Chemically peculiar (CP) stars on the upper main sequence, ranging from early B to early F spectral types, exhibit significant deviations from solar surface abundances, with anomalies manifesting as localized chemical over- or under-abundances across their atmospheres \citep{2003EAS.....9..249L, 2007A&A...471..941L}. These anomalies often involve light elements such as helium (He), which may be strongly depleted, and heavy elements like iron-peak and rare-earth species (e.g., Sr, Cr, Eu), which can exceed solar abundances by several dex. The distribution of these elements is frequently non-uniform, forming chemical spots, belts, or rings modulated by stellar rotation.

CP stars are traditionally grouped into four major subtypes based on their spectral characteristics and magnetic properties \citep{1974ARA&A..12..257P, 1984A&A...138..493M, 2004IAUS..224..443P, 2018CoBAO..65..223G}:  (i) CP1 stars, or Am/Fm stars, exhibit enhanced metallic lines but typically lack strong magnetic fields;  (ii) CP2 stars, also known as Bp/Ap stars, are magnetic and chemically peculiar with pronounced Si, Sr, and Eu over-abundances;  (iii) CP3 stars are HgMn stars, characterized by peculiar mercury and manganese lines; and  (iv) CP4 stars, or He-weak stars, are defined by strong helium under-abundances.

The CP2 and CP4 subclasses, commonly referred to as magnetic chemically peculiar (mCP) stars, are distinguished by the presence of strong, globally organized magnetic fields, often with dipolar topology and field strengths ranging from a few hundred Gauss to several tens of kiloGauss \citep{2007A&A...475.1053A, 2011IAUS..273..249K, 2019MNRAS.490..274S}. These fields are believed to be of fossil origin, i.e., remnants from earlier evolutionary stages or star formation processes \citep{2014IAUS..302..338B}.

The surface abundance patterns observed in CP stars are explained primarily through atomic diffusion in radiative atmospheres \citep{1970ApJ...160..641M, 2005ApJ...625..548R}. In the absence of significant turbulent mixing or convection, and under the stabilizing influence of a magnetic field, radiative levitation can elevate certain elements to the photosphere, while gravitational settling pulls others downward. The efficiency of these processes depends sensitively on atomic properties, local temperature, and magnetic field geometry. The magnetic field suppresses convective motions and horizontal diffusion, leading to the formation of stable chemical spots aligned with magnetic field structures \citep{2015ads..book.....M, 2019EAS....82..345A}.

The observational consequences of this inhomogeneity are profound. As CP stars rotate, the chemical spots traverse the visible disk, producing rotational modulation in brightness, spectral line strengths, and longitudinal magnetic field measurements. This phenomenon is well explained by the oblique rotator model \citep{1949Obs....69..191B, 1950Natur.165..195S}, wherein the magnetic axis is inclined  to the rotational axis. Stars exhibiting such periodic variability are classified  {as} $\alpha^2\, Canum Venaticorum$ 
 (ACV) variables. These variations are typically stable over long timescales, making ACV stars ideal targets for rotational studies and magnetic field mapping through Doppler imaging and Zeeman Doppler imaging techniques \citep{2010A&A...524A..66S, 2012A&A...542A.116K}.

In addition to the rotational modulation, a set of mCP stars pulsates in high‑overtones $p$‑modes (6--24\,min), offering unique asteroseismic insight into the stellar interiors \citep{2006Kurtz, 2012MNRAS.420..283S, 2015JApA...36...33J, 2021MNRAS.506.1073H, 2022ARA&A..60...31K}. The co-existence of rotation and pulsations allow for the investigation of various physical process, such as transport of the angular momentum. In the past, such candidates have been searched and studied through systematics surveys (e.g., \citep{1991MNRAS.250..666M}). In the context of the present work, one of the longest such ground-based surveys is the Nainital--Cape (N--C) Survey  conducted between the Aryabhatta Research Institute of Observational Sciences (ARIES) and South African Astronomical Observatory (SAAO), South Africa, where several asteroseismic candidates were discovered \citep{2001AA...371.1048M, 2003MNRAS.344..431J, 2006A&A...455..303J, 2009A&A...507.1763J, 2010MNRAS.401.1299J, 2012MNRAS.424.2002J, 2016A&A...590A.116J, 2017MNRAS.467..633J, 2022MNRAS.510.5854J}.

Modern space-based observatories such  {as} 
 \kepler, \textit{Transiting Exoplanet Survey Satellite} (\tess), and \brite\ have revolutionized the study of CP stars by enabling high-precision, uninterrupted time-series photometry, uncovering new variability classes, rotational modulation, and pulsation signatures in hundreds of mCP candidates \citep{2019A&A...622A..67B, 10.1093/mnras/stad3800, 2023A&A...671A.154B}. Combined with large-scale spectroscopic efforts and magnetic field surveys (e.g., ESPaDOnS, HARPSpol, MiMeS, and BinaMIcS), this has greatly improved statistical constraints on the incidence and properties of magnetic CP stars across different stellar environments and evolutionary stages \citep{2014IAUS..302..265W, 2019MNRAS.490..274S, 2020MNRAS.499.5049S}.

In this study, we investigate the photometric and spectroscopic characteristics of  AL\,Col. Despite limited previous attention, initial classifications suggest it may belong to the CP2 or CP4 subclass, motivating a more detailed analysis. Using new time-series photometry and high-resolution spectroscopy, we aim to characterize its variability, derive its atmospheric parameters and elemental abundances, and assess its position within the framework of mCP star phenomenology. Our goal is to contribute to the understanding of magnetic field-driven chemical stratification and variability in upper main-sequence stars and to place AL\,Col in the broader context of CP star evolution.

The structure of this paper is as follows. Section\,\ref{sample_selection} outlines the selection of the target star. In Section\,\ref{sec:TESS}, we present the analysis of \tess\ photometric data, including variability characterization through wavelet and autocorrelation methods. Section\,\ref{sec:spectroscopy} details the derivation of fundamental stellar parameters and chemical abundances using high-resolution spectral synthesis. The evolutionary status of the star is discussed in Section\,\ref{sec:evolution}. In Section\,\ref{sec:spot}, we analyze the spatial and temporal characteristics of surface chemical spots. Finally, Section\,\ref{Sec:conclusion} summarizes our main findings and discusses future directions for this study.

\section{Sample Selection}
\label{sample_selection}
\citet{2020MNRAS.493.3293B} determined rotational parameters for more than 300 magnetic CP stars by utilizing ground-based archival photometry from surveys such as ASAS-3, KELT, and MASCARA. From this compilation, we selected a southern hemisphere target for which both the high-precision \tess\ light curves and high-resolution spectra with high signal-to-noise ratio (SNR) are available.  
Our selected target is the $\alpha^2$~CVn-type variable star AL\,Col (HD\,46462), classified as an ApSi star. 

\section{\emph{TESS} Photometry}
\label{sec:TESS}

\tess\ was launched on 18 April 2018 with the primary goal of detecting exoplanets via the transit method. The mission observes nearly the entire sky, which is divided into 26 sectors, each covering a field of $24^\circ \times 96^\circ$. Each sector is continuously monitored for approximately 27.4 days using four wide-field cameras. Each camera hosts four $2\mathrm{K} \times 2\mathrm{K}$ CCD detectors, enabling high-precision photometry over a large field of view.

AL\,Col was observed by \tess\ during the observation of sectors 6 and 7. The  {Python-based} 
 \texttt{ {lightkurve}
} package \citep{collaboration2018lightkurve} was employed to search, download, and analyze the \tess\ time-series data for these sectors. We utilized the short-cadence data  (120\,s) processed by the Science Processing Operations Center (SPOC) pipeline.

Although the Pre-search Data Conditioning Simple Aperture Photometry (PDCSAP) flux is generally preferred due to its correction for systematics and long-term instrumental drifts, in the case of AL\,Col, the raw Simple Aperture Photometry (SAP) flux was found to be of higher quality. Outliers were removed using a sigma-clipping algorithm where necessary. The \tess\ flux values (in $\mathrm{e^{-}\,s^{-1}}$ {)} 
 were converted to magnitudes in the \textit{T}-band using the following standard flux-to-magnitude relation:

\begin{equation}
I_{\mathrm{mag}} = -2.5 \log(\mathrm{FLUX}) + 20.44.
\label{eq:fmr}
\end{equation}

\begin{figure}[H]
\includegraphics[width=0.9\textwidth]{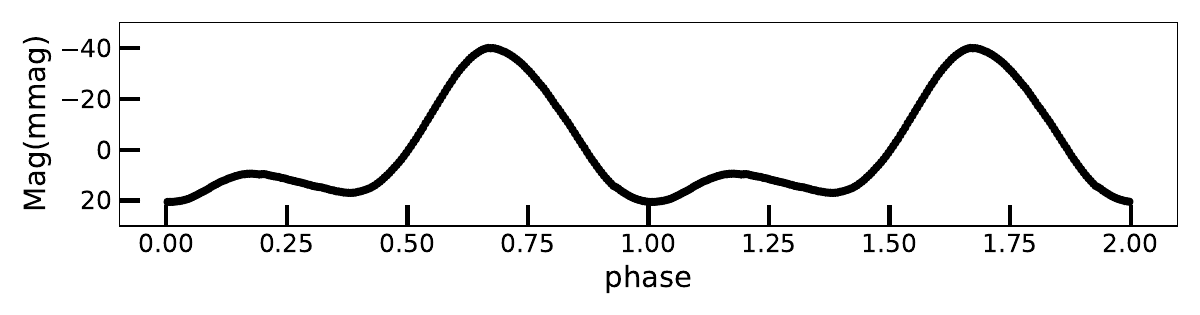}
\caption{ {Phase-folded} 
 light curve of AL\,Col using a reference epoch of $t_0$ = 2,458,471.6706 days (time corresponding to minimum brightness) and a rotation period of 10.35733 d. The data have been binned into 200 phase intervals to improve clarity and reveal the rotational variability.}
\label{fig:fig2}
\end{figure}

To investigate the rotational modulation, the time-series data were phase-folded using the rotation period $\mathrm{P}_{\mathrm{rot}} = 10.35733$ d, following the ephemeris:

\begin{equation}
\phi_{i} = 2\pi \left( \frac{t_{i} - t_{0}}{P_{\mathrm{rot}}} - \left\lfloor \frac{t_{i} - t_{0}}{P_{\mathrm{rot}}} \right\rfloor \right),
\label{eq:phasefold}
\end{equation} \textls[-15]{where $t_i$ is the Barycentric Julian Date (BJD) of each observation and $t_0 = 2,458,471.6706$\,days} is the reference epoch. Figure\,\ref{fig:fig2} shows the resulting phase-folded light curve, binned into 200 phase intervals. The observed light variation exhibits a clear non-sinusoidal shape with a double-peaked structure, indicative of surface brightness inhomogeneities caused by two dominant chemical spots on the stellar surface.

To identify the periodic components of the photometric variability, we performed a Discrete Fourier Transform (DFT) of the combined \tess\ light curve using the \textsc{Period04} software package \citep{2005CoAst.146...53L}, up to the Nyquist frequency ($\nu_{\mathrm{nyq}} \sim 360$\,d$^{-1}$). Significant frequency peaks were identified using the SNR threshold of $\geq$5.2, as recommended by \citet{2021AcA....71..113B}. The dominant frequencies were iteratively prewhitened, and the residuals were reanalysed until no significant frequencies remained.

\begin{figure}[H]
\includegraphics[width=0.90\textwidth]{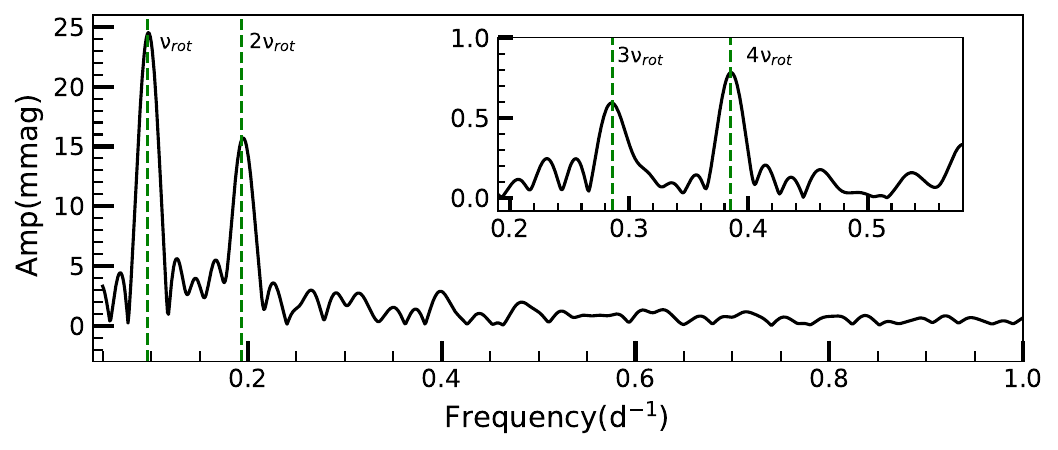}
\caption{Amplitude spectrum of AL Col corresponding to the time-series data in Figure \ref{fig:fig2}. The vertical green lines indicate the observed rotational frequency and its second harmonic. After removing the two dominant frequencies, the inset displays the third and fourth low-amplitude harmonics.} 
\label{fig:fig3}
\end{figure}

Figure\,\ref{fig:fig3} presents the resulting amplitude spectrum of the combined data. The most prominent peak is found at $\nu = 0.09655$\,d$^{-1}$, corresponding to a rotation period of \linebreak  $P_{\mathrm{rot}} = 10.35733$\,d---consistent with the value reported by \citet{2020MNRAS.493.3293B}. The detection of higher-order harmonics further supports the presence of non-uniform surface features modulating the light curve. The identified frequencies, amplitudes, phases, and SNRs are summarized in Table\,\ref{tab:table1}. To confirm the nature of the detected signals, we performed wavelet and autocorrelation analyses of the time-series data.

\vspace{-3pt}
\vspace{-3pt}
\begin{table}[H]
\begin{center}
\caption{ {Listed} 
 are the frequencies, amplitudes, phases, signal-to-noise ratios (SNRs), and identified combination frequencies derived from the Fourier Transform of the \tess\ light curve, constructed by combining time-series data from sectors 6 and 7.} 
\medskip
\begin{tabularx}{\textwidth}{crCRr}
\toprule
\textbf{Frequency}  \boldmath{\textbf{(d$^{-1}$)}} & \textbf{Amplitude (mmag)}& \textbf{Phase}  & \textbf{SNR}    & \textbf{Combination Frequencies} \\
\midrule
0.09655     & 24.2324 	& 0.9878 & 210.93 &  {$ \nu_\mathrm{rot}$} 
 \\
0.19296	    & 14.6962 	& 0.4103 & 137.74 & $2\nu_\mathrm{rot}$\\
0.38499	    & 0.8011 	& 0.7481 & 8.34   & $4\nu_\mathrm{rot}$\\
0.28611	    & 0.6341 	& 0.8373 & 6.28   & $3\nu_\mathrm{rot}$\\
\bottomrule
\end{tabularx}
\label{tab:table1}
\end{center}
\end{table}
\subsection*{Wavelet and Autocorrelation Analysis}

Low-amplitude variations in stellar brightness, such as those caused by eclipses or rotational modulation, occasionally remain undetected through conventional time-series analysis. To enhance the detection sensitivity and examine the temporal evolution of such modulations, advanced time–frequency analysis techniques such as wavelet transforms and autocorrelation functions (ACFs) are frequently employed. These methods are particularly effective in analyzing non-stationary signals and offer robustness against temporal evolution of active regions.

\begin{figure}[H]
\includegraphics[width=0.8\textwidth]{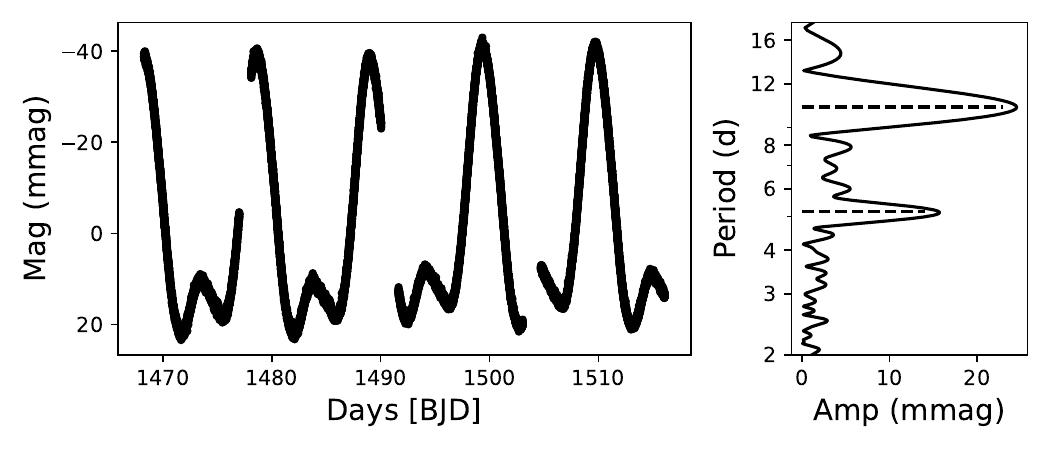}
\includegraphics[width=0.8\textwidth]{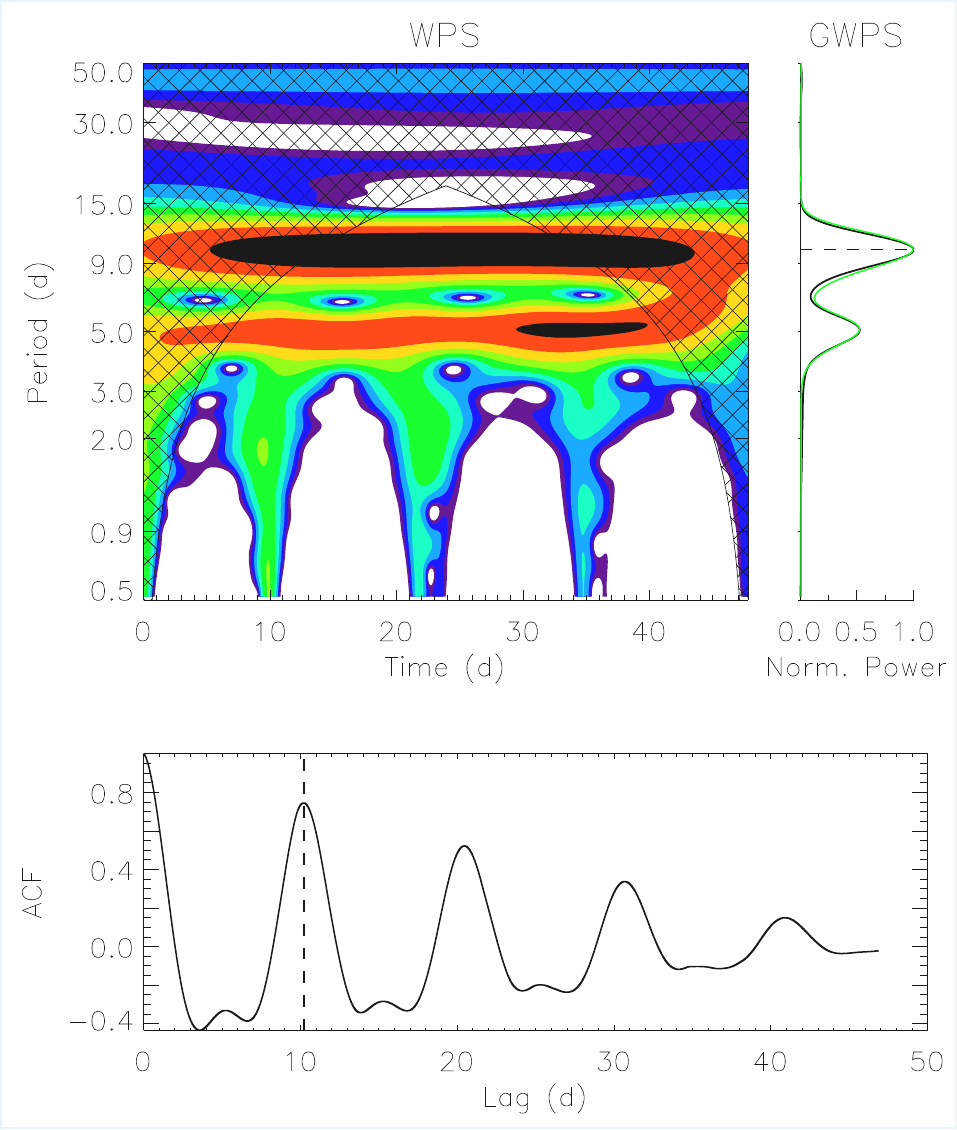}
\caption{{(\textbf{ {Top panels}
})}
: The left sub-panel displays the stellar light curve (time-series photometry), while the right sub-panel presents the corresponding Fourier amplitude spectrum, highlighting dominant periodicities.
{(\textbf{ {Middle panels}})}: The left sub-panel shows the wavelet power spectrum (WPS), where black indicates the highest power and blue represents weaker signals. The red and yellow regions trace strong periodicities, such as stellar rotation and its first harmonic. The shaded region marked by black crosses denotes the cone of influence, where edge effects render the wavelet results unreliable. The right sub-panel shows the Global Wavelet Power Spectrum (GWPS), derived by averaging the WPS over the full timespan of the data.
{(\textbf{ {Bottom panel}})}: The autocorrelation function (ACF) of the light curve is plotted for time lags between 0 and 50 days. Recurrent peaks in the ACF suggest the presence of quasi-periodic variability, often associated with stellar rotation.}
\label{fig:wavelet}
\end{figure}

Wavelet analysis, in particular, is a powerful tool for detecting signatures of rotation, magnetic activity, and pulsations in stellar atmospheres \citep{2014A&A...568A..34B}. For the present study, we utilized a wavelet analysis code jointly developed by Prof. Rafael Garcia and Prof. Savita Mathur. This code employs a Morlet wavelet, a complex sinusoidal function modulated by a Gaussian envelope, to perform convolution with the light curve. Figure\,\ref{fig:wavelet} presents the resulting wavelet power spectrum alongside its corresponding Global Wavelet Power Spectrum (GWPS).

The local wavelet power spectrum captures the distribution of signal energy in both time and frequency domains, while the GWPS is obtained by integrating the local spectrum over time, thereby revealing the average amplitude as a function of frequency. From the GWPS, we identified two dominant periodicities at $10.1043 \pm 1.6809$ d and $5.0521 \pm 0.8411$ d. These values are close to those derived from the Fourier analysis (see Table\,\ref{tab:table1} and Figure\,\ref{fig:fig3}).

To independently validate these periodicities, we also applied the autocorrelation function (ACF) technique to the \tess\ light curve. This method, which evaluates the correlation of the signal with itself at varying time lags, is effective in identifying rotational modulation from co-rotating starspots. The resulting normalized ACF displays a series of quasi-periodic local maxima, indicative of persistent surface features. The first significant peak beyond zero lag yields a dominant autocorrelation period of \mbox{$P_{\mathrm{ACF}} = 10.2118 \pm 1.3074$ d} (see \linebreak  Figure\,\ref{fig:wavelet}), again in agreement with the results from both the wavelet and DFT analyses.

As seen in Figure \ref{fig:wavelet}, the GWPS exhibits two prominent peaks that persist over the full observation baseline. These are represented in the local wavelet power map by warm color gradients (yellow to red), with the dark black regions indicating the strongest power. The primary rotation period maintains a stable amplitude across the observation span, while the secondary period shows a transient behavior---intensifying and subsequently fading after approximately 30 days of observation.

This behavior is characteristic of surface inhomogeneities such as starspots. When two active longitudes are positioned roughly $180^\circ$ apart on opposite hemispheres, the harmonic content of the light curve increases \citep{Mathur}. The dominant active region corresponds to the fundamental rotation period ($\sim$10.1581 d), while the secondary spot gives rise to the  {second} 
 harmonic, observed at $\sim$5.1471 d in both the ACF and GWPS. These findings reinforce the hypothesis that the photometric variability of AL\,Col is driven by rotational modulation due to two major surface chemical spots, which are diametrically opposite.

\section{High-Resolution Spectroscopy}
\label{sec:spectroscopy}
A high-resolution spectrum of AL\,Col was obtained using the High Accuracy Radial Velocity Planet Searcher (HARPS), mounted on the ESO 3.6\,m telescope at La Silla Observatory, Chile. The wavelength-calibrated spectra were retrieved from the European Southern Observatory (ESO) Science  {Portal}
\endnote{\url{https://archive.eso.org/scienceportal/home}} accessed on 8 December 2008. 
 Two exposures, each with an integration time of 300\,s, were acquired on 8 December 2007, yielding a resolving power of $R$$\sim$115,000 and covering the wavelength range 3780--6900\,\AA.

Continuum normalization was performed using a custom Python-based routine, which fits a Chebyshev polynomial to the local maxima representing the spectral upper envelope. To improve the signal-to-noise ratio (SNR), the two spectra were combined using a median-combination technique, resulting in a final SNR of 125.2 at 5500\,\AA.

\subsection{Basic Physical Parameters}
\label{sec:basic_parameters}
Accurate determination of fundamental stellar parameters---such as effective temperature (\teff), surface gravity (\logg), and global metallicity  {([M/H])} 
---is essential for characterizing the star’s atmospheric properties and for constraining asteroseismic and evolutionary models. Spectroscopic techniques remain among the most reliable tools for obtaining these parameters, though initial estimates are typically required to optimize convergence. 

Preliminary estimates of the stellar parameters were derived by comparing the observed SED with synthetic photometric fluxes from theoretical models.

\subsection{Spectral Energy Distribution}
\label{sec:sed}
To estimate the effective temperature, we performed an SED analysis by assembling multi-wavelength photometric fluxes spanning the optical to infrared regimes. The observational SED was constructed using data from various photometric systems: the Strömgren–Crawford  {$uvby$} 
 system, Tycho-2, Johnson UBV, Hipparcos, Gaia DR3, CMC15, 2MASS, WISE, NEOWISE, and DENIS catalogues \citep{2000A&AS..143...23O, 2015A&A...580A..23P, 2000A&A...357..367H, 2022arXiv220606215G, 2016A&A...595A...1G, 2006AJ....131.1163S, 2010AJ....140.1868W, 2000A&AS..141..313F, 1999A&A...349..236E, 2014ApJ...792...30M}. Data retrieval was facilitated through  {VizieR}
\endnote{{\url{http://cdsarc.u-strasbg.fr/viz-bin/Cat?II/168}}, {\url{https://cdsarc.unistra.fr/viz-bin/cat/I/327}}} 
 and the Spanish Virtual Observatory (SVO)\endnote{\url{http://svo2.cab.inta-csic.es/vocats/cmc15/}}.

The observed fluxes were plotted as a function of their respective effective wavelengths, and a synthetic SED was generated by fitting theoretical flux distributions to the multi-band data using an iterative $\chi^2$ minimization technique. The best-fitting model yields an effective temperature of \teff\,$=11{,}750 \pm 125$\,K (see Figure\,\ref{fig:fig5}). The derived \teff\, serves as initial inputs for the detailed spectroscopic analysis discussed in the next section.

\begin{figure}[H]
\includegraphics[width=0.85\textwidth]{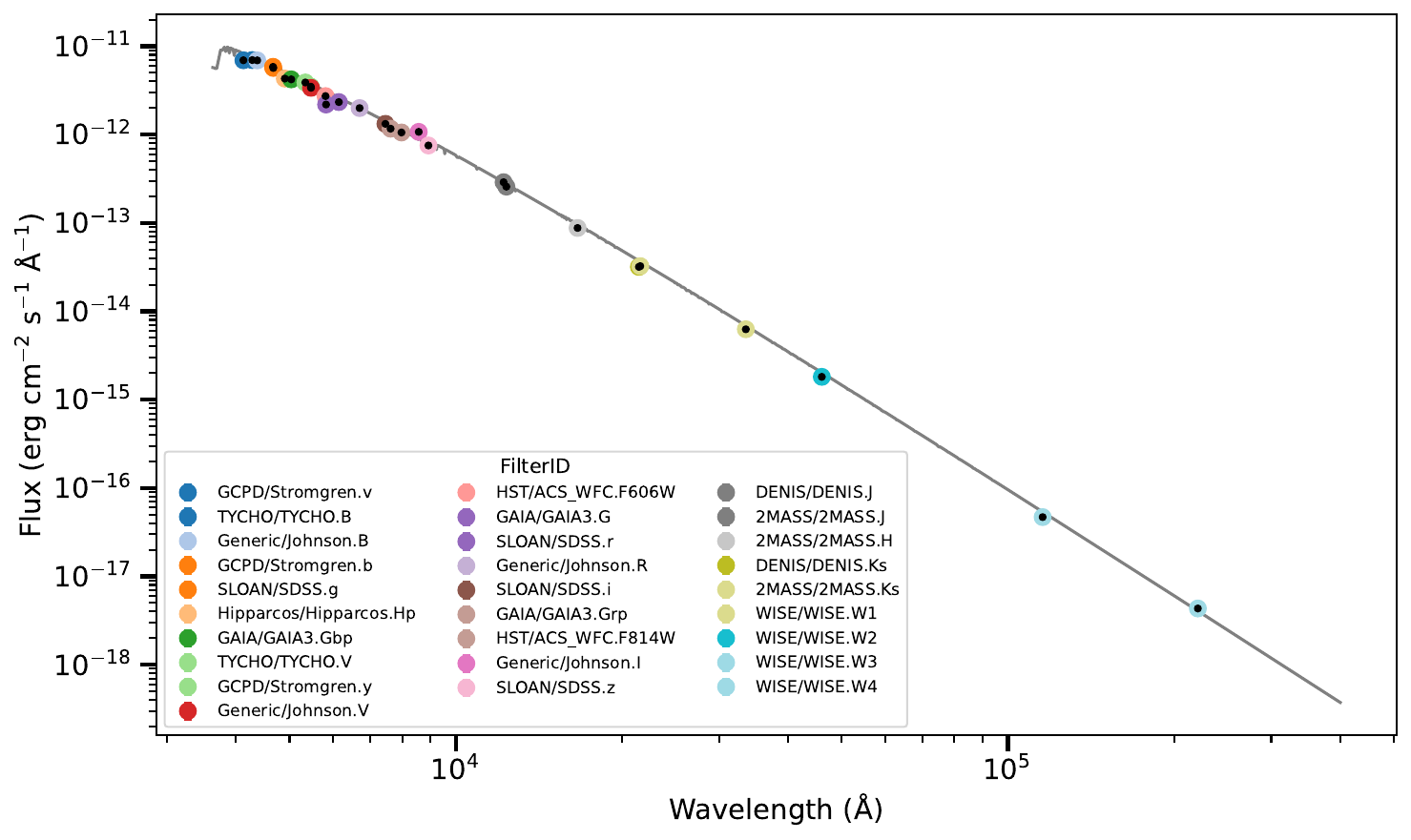}
\caption{Spectral energy distribution (SED) spanning the optical to infrared wavelengths. Colored dots represent observed fluxes in different photometric bands, as indicated in the legend. The solid gray line corresponds to the best-fitting theoretical model (\teff\,=\,11,750\, $\pm$\,125\, K). Both the axes are plotted in logarithmic scale.}
\label{fig:fig5}
\end{figure}

\subsection{Spectral Synthesis with \texttt{pySME}}
\label{sec:sme} 
The atmospheric parameters derived from the SED analysis were used as initial inputs to compute synthetic spectra for detailed spectroscopic modeling. The estimated (\teff) from SED, along with \logg\ = 4.0 dex (assuming the target to be a main-sequence star), were adopted to extract the atomic and molecular line data from the VALD3 database \citep{pakhomov2017vald3currentdevelopments}, which includes isotopic and hyperfine structure components essential for high-resolution modeling.

To align the synthetic spectra with the observed HARPS data, a Doppler correction was applied, yielding a measured radial velocity of $v_{\mathrm{rad}} = 28$\,km\,s$^{-1}$. Grids of synthetic spectra were generated using the Python-based package \texttt{pySME} \citep{Wehrhahn_2023}, which implements the \texttt{Spectroscopy Made Easy} (SME) framework. This code adopts plane-parallel model atmospheres from ATLAS9 \citep{castelli2004newgridsatlas9model} under the assumption of local thermodynamic equilibrium (LTE). SME determines atmospheric parameters and chemical abundances by fitting synthetic spectra to selected regions of the observed spectrum. A $\chi^2$ minimization approach is employed to iteratively adjust the parameters until the best match is achieved.

A line mask was constructed by selecting diagnostic lines that are sensitive to variations in different atmospheric parameters. The fitting process searched for the optimal combination of \teff, \logg, and projected rotational velocity (\vsini) by minimizing residuals between the synthetic and observed spectra.

For the iterative fitting procedure, the initial parameters were set as follows: \teff\ from the SED analysis, \logg\ was fixed at 4.0 (typical of main-sequence stars), solar metallicity ([M/H] = 0), $v_{mic}=2$\,\kms\ , and \vsini\ was estimated by visually matching the width of metal lines. Since Balmer lines are particularly sensitive to \teff\ and \logg, we refined these parameters by fitting synthetic profiles to the observed H$\beta$ (4861\,\AA) and H$\alpha$ (6563\,\AA) line profiles. The projected rotational velocity \vsini\ was determined by fitting the Mg~\textsc{i} triplet region (5160--5187\,\AA). Figure~\ref{fig:spectra} shows the comparison between the observed and best-fit synthetic spectra in the regions around the H$\beta$ and H$\alpha$ lines.

\begin{figure}[H]

\includegraphics[width=1\textwidth]{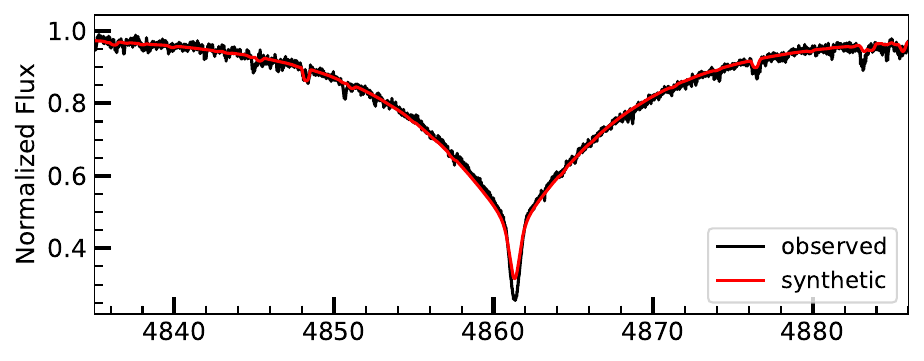}

\caption{\textit{Cont.}}
\end{figure}
\begin{figure}[H]\ContinuedFloat

\includegraphics[width=1\textwidth]{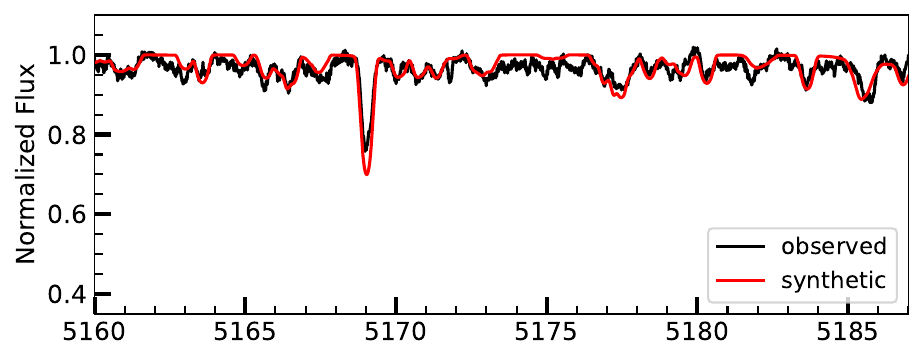}
\includegraphics[width=1\textwidth]{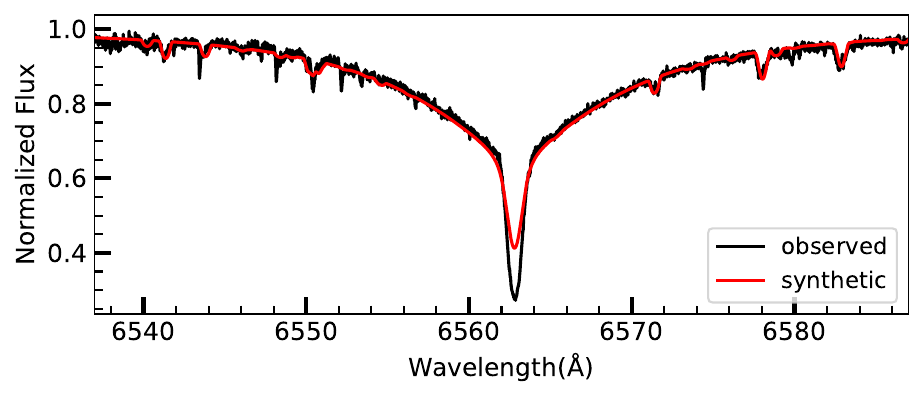}
\caption{ {Comparison} 
 between the observed (black) and synthetic (red) spectra in three key regions: the H$\beta$ (\textbf{top panel}), the Mg\,{\sc i} triplet region (\textbf{middle panel}), and the H$\alpha$ (\textbf{bottom panel}). The synthetic spectra were computed using the best-fitting atmospheric parameters.}
\label{fig:spectra}
\end{figure}

Parameter uncertainties were estimated from the standard deviation of the residuals between the observed and synthetic spectra within each fitted region. The final atmospheric parameters are \teff  = 13,814 $\pm$ 400 K, \logg = 4.09 $\pm$ 0.08\,dex, and \vsini = 16 $\pm$\,1\,\kms.

\subsection{Photospheric Chemical Abundance Analysis}
\label{Sec:abundances}

Chemical abundances provide valuable insights into the patterns of elemental enhancement and depletion in stellar atmospheres, which in turn inform our understanding of stellar structure, evolution, and diffusion processes. We performed a detailed analysis of the photospheric composition of AL\,Col under the assumption of local thermodynamic equilibrium (LTE). Individual elemental abundances were derived by fitting synthetic line profiles to the observed HARPS spectrum using the \texttt{BinMag6} visualization tool\endnote{\url{http://www.astro.uu.se/~oleg/binmag.html}} \citep{kochukhov2018binmag}. The atmospheric parameters determined in Section~\ref{sec:sme} (\teff, \logg, and \vsini) were held fixed throughout the abundance analysis.

A total of 25 chemical elements, ranging from helium (He) to europium (Eu), were investigated. Both neutral and ionized species (first and second ions) were considered based on the presence of unblended or mildly blended absorption lines in the optical spectral range. The synthetic spectra used for fitting were interpolated from the model grid developed by \citet{Tkachenko_2015}, which includes 41,888 LTE models spanning a metallicity range of [M/H] = $-0.8$ to $+0.8$\,dex (in steps of 0.1\,dex) and assumes a fixed microturbulent velocity of $v_{\mathrm{mic}} = 2$\,km\,s$^{-1}$. The grid offers high resolution in \teff\ (100\,K steps between 4500--10,000\,K and 250\,K steps between 10,000--22,000\,K) and in \logg\ (2.5--5.0\,dex for low \teff, 3.0--5.0\,dex for high \teff).

For AL\,Col, we selected the model atmosphere closest to the spectroscopically determined parameters. Atomic line data were extracted from the VALD3 database, and relevant absorption features were identified in the 4500--6850\,\AA\ wavelength range. For each element, spectral sub-regions containing suitable diagnostic lines were selected for abundance determination. Each sub-region yielded an independent abundance measurement, and the final elemental abundance was computed as the mean of the values derived from multiple segments.

\begin{figure}[H]
\includegraphics[angle=180,width=0.9\textwidth]{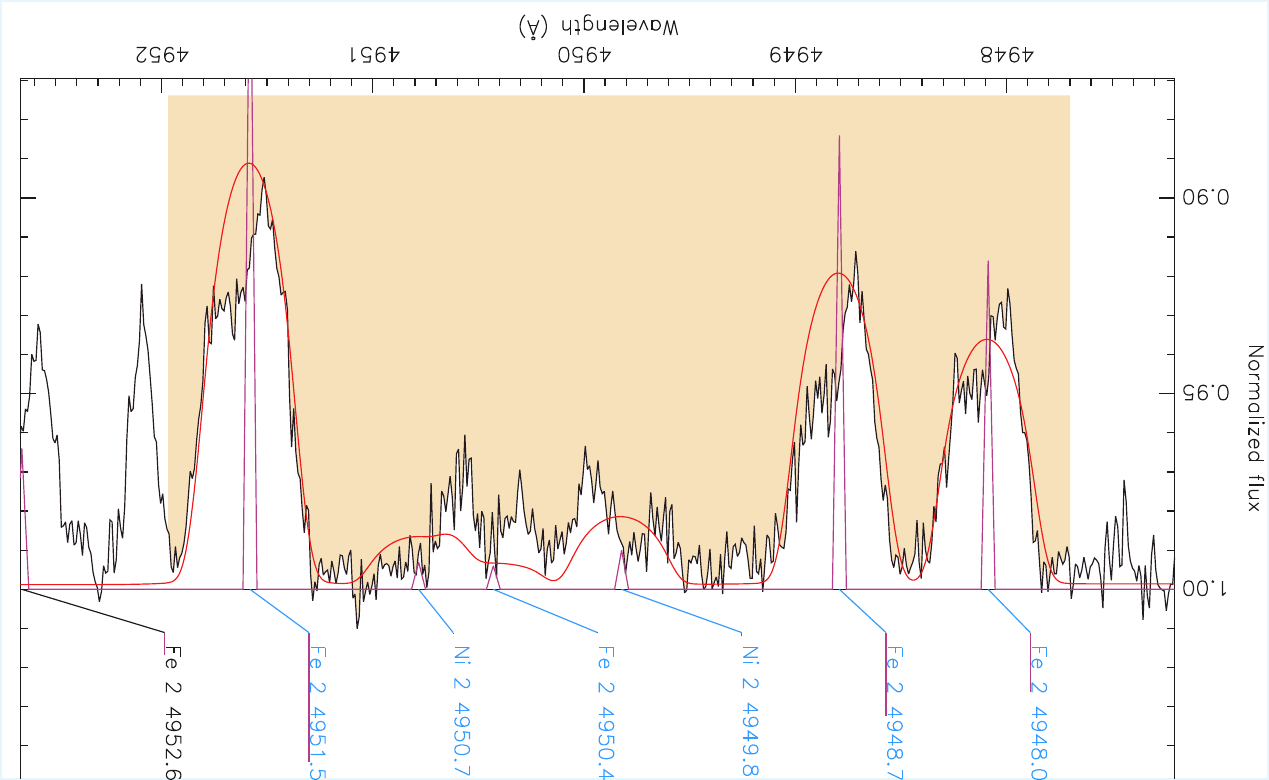}
\caption{ {A} 
 selected piece of the spectrum in a narrow range if
 the wavelength range is 4943 \AA--4953 \AA\, for the clear visibility. Lines of various ions are marked using {\it  {BinMag6} 
} interface. The observed and synthetic spectra are shown with black and red colors, respectively. The shaded region is the active region; i.e., 
only lines under that region are fitted (e.g., fitted Fe\,{\sc ii} lines.)}
\label{fig:fig6}

\end{figure}

 {The} 
 interface of \texttt{BinMag6} is displayed in Figure \ref{fig:fig6}, illustrating the spectral window spanning approximately 4947--4953 \AA. The shaded region indicates the active fitting area, where spectral lines corresponding to the elements present have been modeled. In this analysis, emphasis was placed on the strongest lines, particularly three Fe\,\textsc{ii} transitions. Spectral features associated with Ni\,\textsc{ii} and a weaker Fe\,\textsc{ii} line within this interval exhibited an insufficient signal-to-noise ratio and were therefore excluded from the analysis. A notable feature is the presence of a central bump in the cores of the Fe\,\textsc{ii} lines, a phenomenon commonly observed in metallic lines (Cr, Ni, Nd, Pr, Eu, etc.) of magnetic chemically peculiar (mCP) stars. Additionally, asymmetries---particularly bumps on line cores---are attributed to line profile variations induced by chemical spots on the stellar surface. Accurate modeling of such line profile variations necessitates time-series spectroscopic observations \citep{10.1093/mnras/staf1247}.

\begin{figure}[H]
\includegraphics[width=0.9\textwidth]{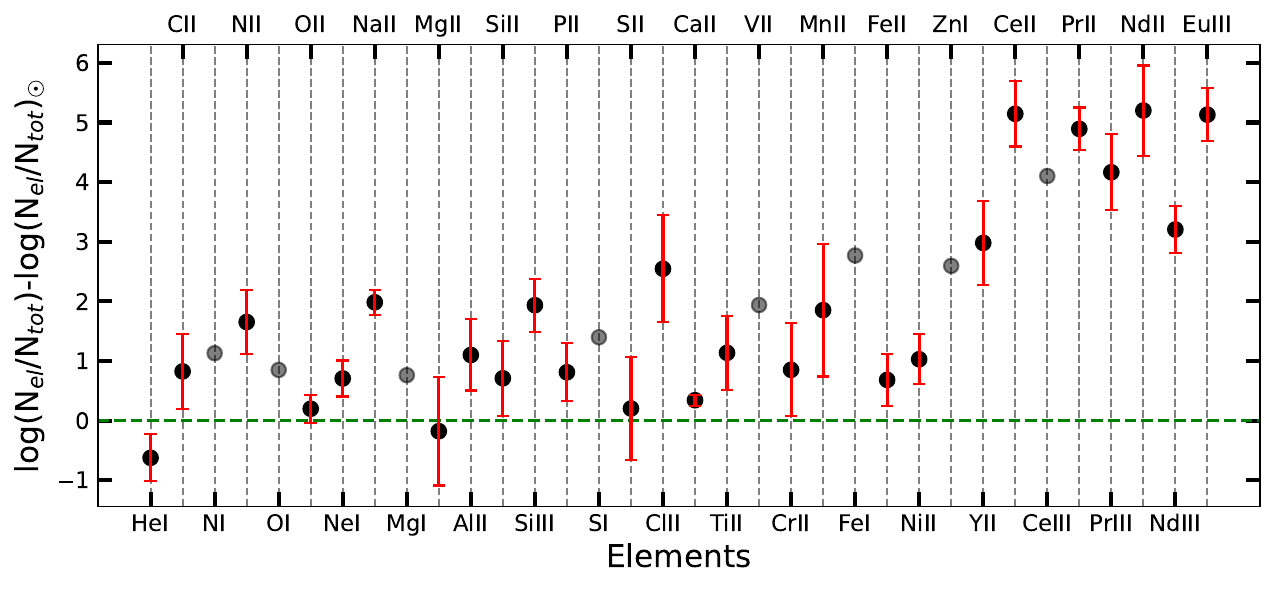}
\caption{ {Graphical} 
 representation of the elemental abundances listed in Table\,\ref{tab:table_abundances}. The horizontal green line indicates the solar abundances from \citet{Asplund_2021}. Black dots represent the mean abundances of elements for which multiple spectral lines were measured; the associated red error bars denote the standard deviation of these measurements. Gray dots indicate elements for which only a single line was available, and these values should be interpreted with caution.}
\label{fig:fig9}
\end{figure}

Elemental abundances are reported in the standard logarithmic notation:
\[
\log \epsilon = \log\left(\frac{N_{\mathrm{el}}}{N_{\mathrm{tot}}}\right) + 12.04,
\]
where \(N_{\mathrm{el}}\) and \(N_{\mathrm{tot}}\) represent the number densities of the element and the total number of atoms, respectively. The final abundances are summarized in Table\,\ref{tab:table_abundances} and visualized in Figure\,\ref{fig:fig9}.

\vspace{-6pt}
\begin{table}[H]
\footnotesize
\caption{  {Chemical} 
 elements analyzed in AL Col, along with their atomic numbers, number of spectral lines used, and mean photospheric abundances. The standard errors of the mean abundances, calculated as the standard deviation of individual line measurements, are provided. Solar photospheric abundances adopted from \citet{Asplund_2021} are listed for comparison.} 
\label{tab:table_abundances}
\begin{tabularx}{\textwidth}{LcccCC}
\toprule
\textbf{Element} & \textbf{Atomic No. Z}& \textbf{No. of Lines Used}&\boldmath{\textbf{log $\epsilon_{\rm AL Col}$  {(dex)} 
}}&\textbf{Error (dex)}&\boldmath{\textbf{log $\epsilon_\odot$ (dex)}}\\
\midrule\
He\,{\sc i}& 2& 6& $-$1.71& 0.39& $-$1.09\\
C\,{\sc ii}& 6& 5& $-$2.72& 0.62& $-$3.54\\
N\,{\sc i}& 7& 1& $-$3.04& --& $-$4.17\\
N\,{\sc ii}& 7& 5& $-$2.52& 0.54& $-$4.17\\
O\,{\sc i}& 8& 1& $-$2.46& --& $-$3.31\\
O\,{\sc ii}& 8& 3& $-$3.11& 0.24& $-$3.31\\
Ne\,{\sc i}& 10& 10& $-$3.23& 0.30& $-$3.94\\
Na\,{\sc ii}& 11& 2& $-$3.80& 0.21& $-$5.78\\
Mg\,{\sc i}& 12& 1& $-$3.69& --& $-$4.45\\
Mg\,{\sc ii}& 12& 5& $-$4.63& 0.91& $-$4.45\\
Al\,{\sc ii}& 13& 3& $-$4.47& 0.60& $-$5.57\\
Si\,{\sc ii}& 14& 29& $-$3.78& 0.63& $-$4.49\\
Si\,{\sc iii}& 14& 2& $-$2.55& 0.45& $-$4.49\\
P\,{\sc ii}& 15& 6& $-$5.78& 0.48& $-$6.59\\
S\,{\sc i}& 16& 1& $-$3.48& --& $-$4.88\\
S\,{\sc ii}& 16& 20& $-$4.68& 0.86& $-$4.88\\
Cl\,{\sc ii}& 17& 3& $-$4.14& 0.90& $-$6.69\\
Ca\,{\sc ii}& 20& 2& $-$5.36& 0.09& $-$5.70\\
Ti\,{\sc ii}& 22& 9& $-$5.89& 0.62& $-$7.03\\
V\,{\sc ii}& 23& 1& $-$6.16& --& $-$8.10\\
Cr\,{\sc ii}& 24& 12& $-$5.53& 0.78& $-$6.38\\
Mn\,{\sc ii}& 25& 2& $-$4.73& 1.11& $-$6.58\\
Fe\,{\sc i}& 26& 1& $-$1.77& --& $-$4.54\\
Fe\,{\sc ii}& 26& 133& $-$3.86& 0.44& $-$4.54\\
Ni\,{\sc ii}& 28& 10& $-$4.77& 0.42& $-$5.80\\
Zn\,{\sc i}& 30& 1& $-$4.84& --& $-$7.44\\
Y\,{\sc ii}& 39& 4& $-$6.81& 0.71& $-$9.79\\
Ce\,{\sc ii}& 58& 15& $-$5.27& 0.55& $-$10.42\\
Ce\,{\sc iii}& 58& 1& $-$6.32& --& $-$10.42\\
Pr\,{\sc ii}& 59& 3& $-$6.36& 0.36& $-$11.25\\
Pr\,{\sc iii}& 59& 26& $-$7.08& 0.64& $-$11.25\\
Nd\,{\sc ii}& 60& 4& $-$5.38& 0.76& $-$10.58\\
Nd\,{\sc iii}& 60& 28& $-$7.37& 0.39& $-$10.58\\
Eu\,{\sc iii}& 63& 3& $-$6.35& 0.45& $-$11.48\\
\bottomrule

\end{tabularx}    
\end{table}

The large discrepancies between abundances derived from successive ionization stages, in particular the 1.23\,dex offset between Si\,\textsc{ii} and Si\,\textsc{iii} and the 2.09\,dex offset between Fe\,\textsc{i} and Fe\,\textsc{ii}, strongly point to departures from a simple LTE, homogeneous atmospheric model. In A‑ and B‑type chemically peculiar stars, overionization of minority species under non‑LTE conditions leads to weakened neutral‑stage lines and hence to underestimated LTE abundances \citep{Mashonkina2011,Przybilla2001}. At the same time, vertical stratification of elements, observationally demonstrated in Ap stars, can concentrate both Fe and Si at different optical depths where Fe\,\textsc{i}, Fe\,\textsc{ii}, Si\,\textsc{ii}, and Si\,\textsc{iii} lines form, amplifying ionization imbalances \citep{Ryabchikova2005}. Indeed, Romanovskaya et al.\ (2024) find a $\approx$1.5\,dex abundance jump between the layers where Fe\,{\sc ii-iii} and Si\,{\sc ii-iii} lines form in the Ap star BD\,+00$^{\circ}$\,1659, further underscoring the combined role of stratification and non‑LTE effects \citep{Romanovskaya2024}. Depth‑dependent micro- and macro-turbulences further distort line profiles in opposite senses for neutral versus ionized species \citep{Kochukhov2007}. To reconcile these offsets and to derive a physically consistent abundance profile for AL\,Col, full non‑LTE spectrum synthesis with a stratified abundance model is required (e.g.,\  using TLUSTY/SYNSPEC \citep{Hubeny1995} or DETAIL/SURFACE \citep{Butler1985}), which will also shed light on the operation of diffusion and weak stellar winds in its atmosphere.


\section{Evolutionary Status}
\label{sec:evolution}


Determining the evolutionary status of the target star is crucial for placing its physical and chemical properties in the broader context of stellar structure and evolution. It provides insight into the star’s internal processes, surface composition, and variability mechanisms and is essential for accurately modeling its pulsations and rotational behavior.

To assess the evolutionary status of AL\,Col, we utilized the PAdova and TRieste Stellar Evolution Code (PARSEC; 	Bressan et al. \citep{2012MNRAS.427..127B}) to compute stellar evolutionary tracks and isochrones. The models were generated assuming a solar metallicity of $Z = 0.0152$ \citep{2000ApJ...539..352H}. A grid of evolutionary tracks was computed for a range of stellar masses, and isochrones corresponding to various ages were overlaid on the \teff--\logg\ diagram.

\subsection{Luminosity Estimate}
\label{sec:luminosity}

The stellar luminosity was calculated using the following relation:
\begin{equation}
\log\, \left(\frac{L}{{\rm L}_{\odot}}\right) = -0.4 \left(M_{\rm V} + BC - M_{\mathrm{bol},\odot}\right),
\label{eq:L/L0}
\end{equation} where $M_{\mathrm{bol},\odot} = 4.74$ is the solar bolometric magnitude, and $BC$ is the bolometric correction derived using the empirical relation from \citet{2010torres}. The absolute magnitude $M_V$ in the $V$ band was computed using

\begin{equation}
M_{\rm V} = m_{\rm V} + 5 - 5\,\log{(d)} - A_{\rm V},
\label{eq:Mv}
\end{equation} where $m_{\rm V} = 7.56$ is the apparent visual magnitude, $d = 243$\,pc is the distance estimated from \textit{Gaia}\,DR3 parallax \citep{2021A&A...649A...1G}, and $A_{\rm V} = 1.61$ is the extinction in the $V$ band, derived using the reddening value $E{\rm (B-V)} = 0.52$ \citep{2015MNRAS.451.1396M} and adopting $R_V = 3.1$.

\textls[-15]{Substituting the derived values into Equations\,(\ref{eq:L/L0}) and (\ref{eq:Mv}) yields $\log{(L/{\rm L}_{\odot})} = 2.69 \pm 0.09$,}
which corresponds to a mid-late B-type to early A-type main‐sequence star, in excellent agreement with the spectroscopic classification presented in Section\,\ref{sec:sme}.


\subsection{Radius, Mass, and Age}

The stellar radius was estimated using the following relation:
\begin{equation}
\log\left(\frac{R}{{\rm R}_{\odot}}\right) = 0.5 \log \left(\frac{ L}{{\rm L}_{\odot}}\right) - 2\log \left(\frac{T}{{\rm T}_{\odot}}\right),
\label{eq:R}
\end{equation}
\noindent
yielding a value of \( R = 3.74 \pm 0.48\,{\rm R}_\odot \) that is approximately a solar radius larger than the published value ($2.44\,\rm R_{\odot}$) by \citet{2019AstBu..74...66G}. The underestimated stellar radius in previous studies can primarily be attributed to the use of Hipparcos-based parallaxes and the adoption of luminosities uncorrected for interstellar extinction \citep{2019AstBu..74...66G}. By combining the stellar radius ($R$) and \vsini\ derived in our analysis, we infer an inclination angle of approximately 60$^{\circ}$.

\begin{figure}[H]
\includegraphics[width=1\textwidth]{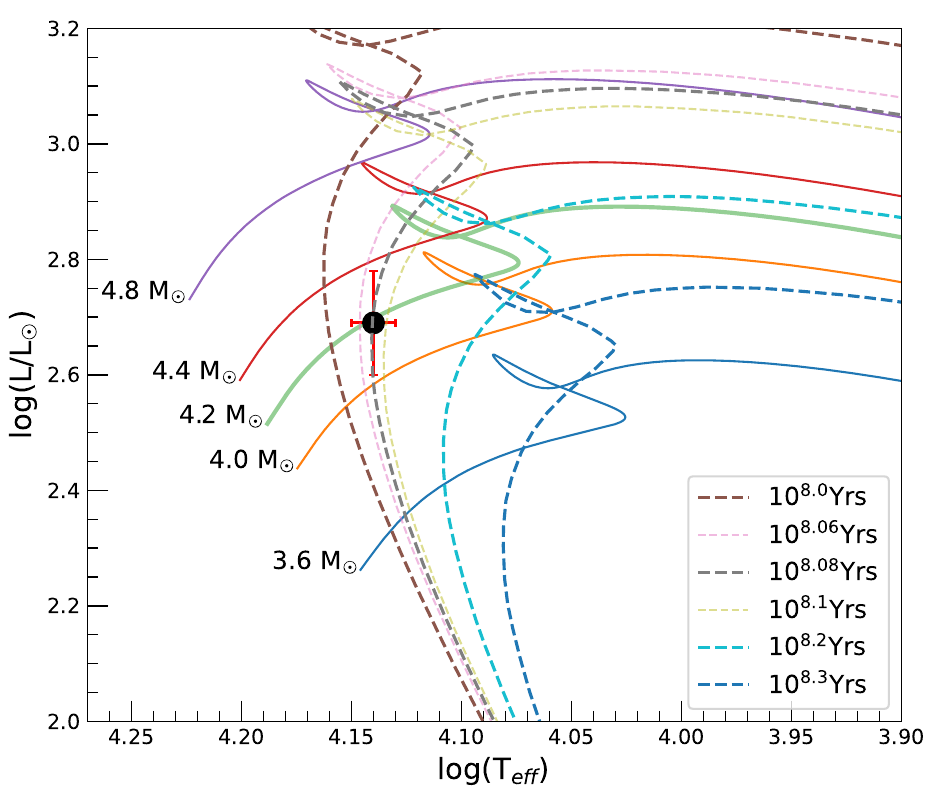}
\caption{  {Location} 
 of AL\,Col in the Hertzsprung--Russell (H--R) diagram is marked with a filled black circle. Vertical and horizontal red lines are representing the errors in luminosity and temperature, respectively. Solid lines represent PARSEC evolutionary tracks for stellar masses ranging from 3.6 to 4.6\,M$_\odot$, while dashed lines show PARSEC isochrones of different ages, as indicated in the legend.}
\label{fig:HRdia}
\end{figure}

The location of the star in the Hertzsprung--Russell (H--R) diagram is illustrated in Figure\,\ref{fig:HRdia}, where evolutionary tracks corresponding to stellar masses ranging from 3.2 to 4.8\,M$_\odot$, in increments of 0.4\,M$_\odot$, are overplotted. The position of the target aligns closely with the evolutionary track of 4.2\,$\pm$\,0.2\,M$_\odot$, consistent with the mass estimate of \( M_\star = 3.9\,{\rm M}_\odot \) reported by \citet{2019AstBu..74...66G}.

By comparing the position of the star on the H--R diagram with theoretical isochrones, we estimate its age to be \(0.120 \pm 0.01\,\mathrm{Gyr}\). This value is approximately twice that reported by \citet{2019AstBu..74...66G} (\(\sim0.055\,\mathrm{Gyr}\)), likely due to an underestimation of stellar luminosity in their study. The diagram suggests that the star is situated at the middle of ZAMS and TAMS, slowly approaching the turn-off point, consistent with its derived physical parameters.

A summary of the fundamental parameters of HD\,46462, including values obtained in this study and from the literature, is presented in Table\,\ref{tab:tab2}.

\begin{table}[H]
\setstretch{1.15}
\caption{  {Basic} 
 parameters of AL\,Col compiled from both the literature and this study. The reference numbers correspond to the following sources:
 $^{1}$ (\citet{2023ApJS..266...11B}), $^{2}$ (\citet{2020MNRAS.493.3293B}), \linebreak  $^{3}$ (\citet{2019AstBu..74...66G}), $^{4}$ (SIMBAD astronomical Database), $^{5}$ (\citet{2015MNRAS.451.1396M}), $^{6}$ (\citet{2021A&A...649A...1G}), $^{7}$ (\citet{2007A&A...474..653V}).}
\label{tab:tab2}
\fontsize{9}{10}\selectfont
\begin{tabularx}{\textwidth}{l C | l C}
\noalign{\hrule height 1.0pt}
\multicolumn{2}{c|}{\textbf{Literature}} & \multicolumn{2}{c}{\textbf{Present Study}} \\
 ine
\textbf{Parameter}     & \textbf{Value}            & \textbf{Parameter} & \textbf{Value} \\
 ine
$T_{\rm eff}$ (K) $^{1}$             & 12,600 $\pm$ 126           & $T_{\rm eff}$ (K) (Spectroscopy)  & 13,814 $\pm$ 400 \\
                                    &                         & $T_{\rm eff}$ (K) (SED)           & 11,750 $\pm$ 125\\
$\log g$ (dex) $^{1}$            & 3.99 $\pm$ 0.03           & $\log g$ (dex)                & 4.09 $\pm$ 0.08 \\
$v \sin i$ (km\,s$^{-1}$) $^{1}$  & 19.49 $\pm$ 0.74          & $v \sin i$ (km s$^{-1}$)      & 16 $\pm$ 1 \\
$P_{\rm rot}$ (d) $^{2}$          & 10.363                    & $P_{\rm rot}$ (d)              & 10.35733 \\
$\log (L/\rm L_{\odot})$ $^3$               & 2.336         & $\log (L/\rm L_{\odot})$                      & 2.69 $\pm$ 0.09\\
$R/\rm R_{\odot}$ $^3$                   & 2.44                    & $R/\rm R_\odot$                   & 3.74 $\pm$ 0.48 \\
$M/\rm M_{\odot}$ $^3$                   & 3.9                      &  $M/\rm M_\odot$                          & 4.2 $\pm$ 0.2 \\
Age (Gyrs) $^3$                      & 0.055                      &   Age (Gyrs)                 &  0.12 $\pm$ 0.01 \\ 
$m_{\rm v}$ (mag) $^{4}$               & 7.56          &            &    \\
B-V $^{4}$                     & $-$0.1         &                    &  \\
$E{\rm (B-V)}$ $^{5}$                    & 0.52                      &                               & \\
plx(mas) $^{6,7}$                     & 4.11 $\pm$ 0.38 (d = 243 pc)             &                               & \\
                                & 2.93 $\pm$ 0.52 (d = 341 pc)             &
                  &\\ 
 ine
\end{tabularx}
\end{table}

\section{Spot Modeling}
\label{sec:spot}

Modeling a stellar surface provides crucial insights into surface dynamics and magnetic activity, as starspots co-rotate with the stellar surface and induce brightness modulations modulated by stellar rotation. We reconstructed the \tess\, light curves from sectors 6 and 7 using the BestrAndom StarSpots Model cAlculatioN (BASSMAN;~\citep{Bicz2022ApJ...935..102B})\endnote{\url{https://github.com/KBicz/BASSMAN}} program to perform inverse modeling of starspots and generate stellar surface maps. The required input parameters for BASSMAN include the star’s rotation period, \teff, radius, and inclination angle (or \vsini). Using a Markov Chain Monte Carlo (MCMC) approach, the program determines spot parameters such as amplitude, size, latitude, and longitude. A three-spot model was adopted, as it provided a better fit to the data than a two-spot configuration.

Once the input parameters are provided, BASSMAN proceeds by positioning the spots on the stellar surface and fitting a model light curve to the observed data. This is accomplished using optimization routines from the scipy library, which iteratively adjust the spot parameters to improve the match between the modeled and observed light curves. The fitting performance is evaluated using correlation coefficients or a reduced $\chi^2$ statistic to determine the model’s reliability. If the fit does not converge or yield satisfactory agreement, the algorithm samples a new set of spot parameters and restarts the fitting process. This approach allows BASSMAN to successfully model light curves affected by both magnetic and chemical spots, ultimately providing best-fit parameters that describe the observed brightness variations with high fidelity.

Our results indicate the presence of three cool surface regions. For Sectors 6 and 7, the model yielded a reduced $\chi^2$ of 4.42 and 5.89, along with spot coverage of 1.49 + 1.99 + 1.00 =\,4.48\% and 1.00 + 2.49 + 1.39 =\,4.88\% which are consistent with analytically derived spot sizes 5.77\% and 5.68\%, respectively, using equations \citep{2013ApJ...771..127N, 2013PASJ...65...49S,2019ApJ...876...58N}:

\begin{equation}
   T_{spot} = 0.751\,T_{\rm eff} -3.58\,\times\,10^{-5}T_{\rm eff}^2 + 808
\end{equation}
 {and} 
\begin{equation}
   \frac{A_{\rm spot}}{A_{\rm star}}\,=\,100\%\,\times\,\frac{\Delta F}{F}\left\lfloor1-\left(\frac{T_{\rm spot}}{T_{\rm eff}}\right)\right\rfloor^{-1}.
\end{equation}

\begin{figure}[H]

\includegraphics[width=0.495\textwidth]{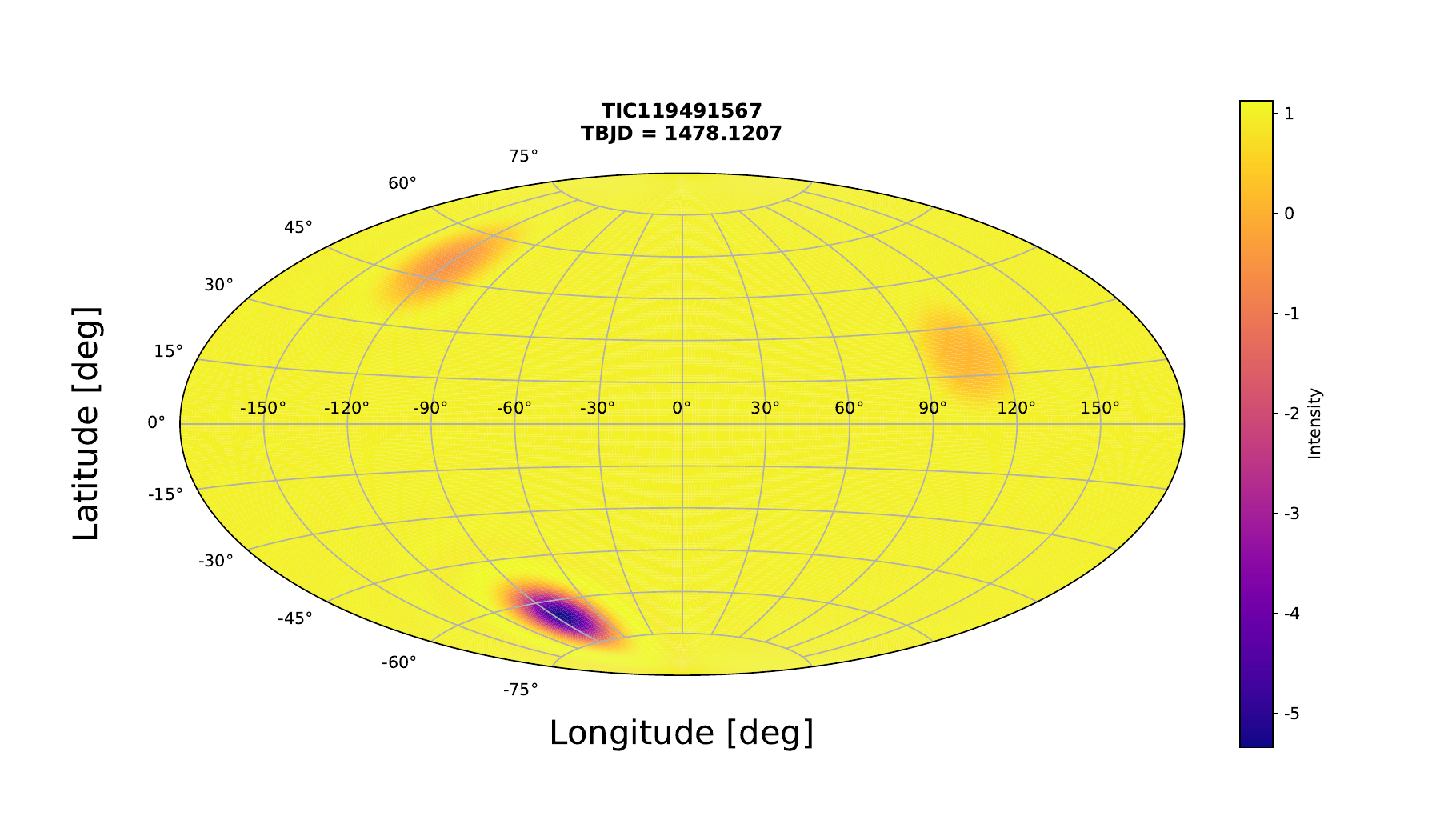}
\includegraphics[width=0.495\textwidth]{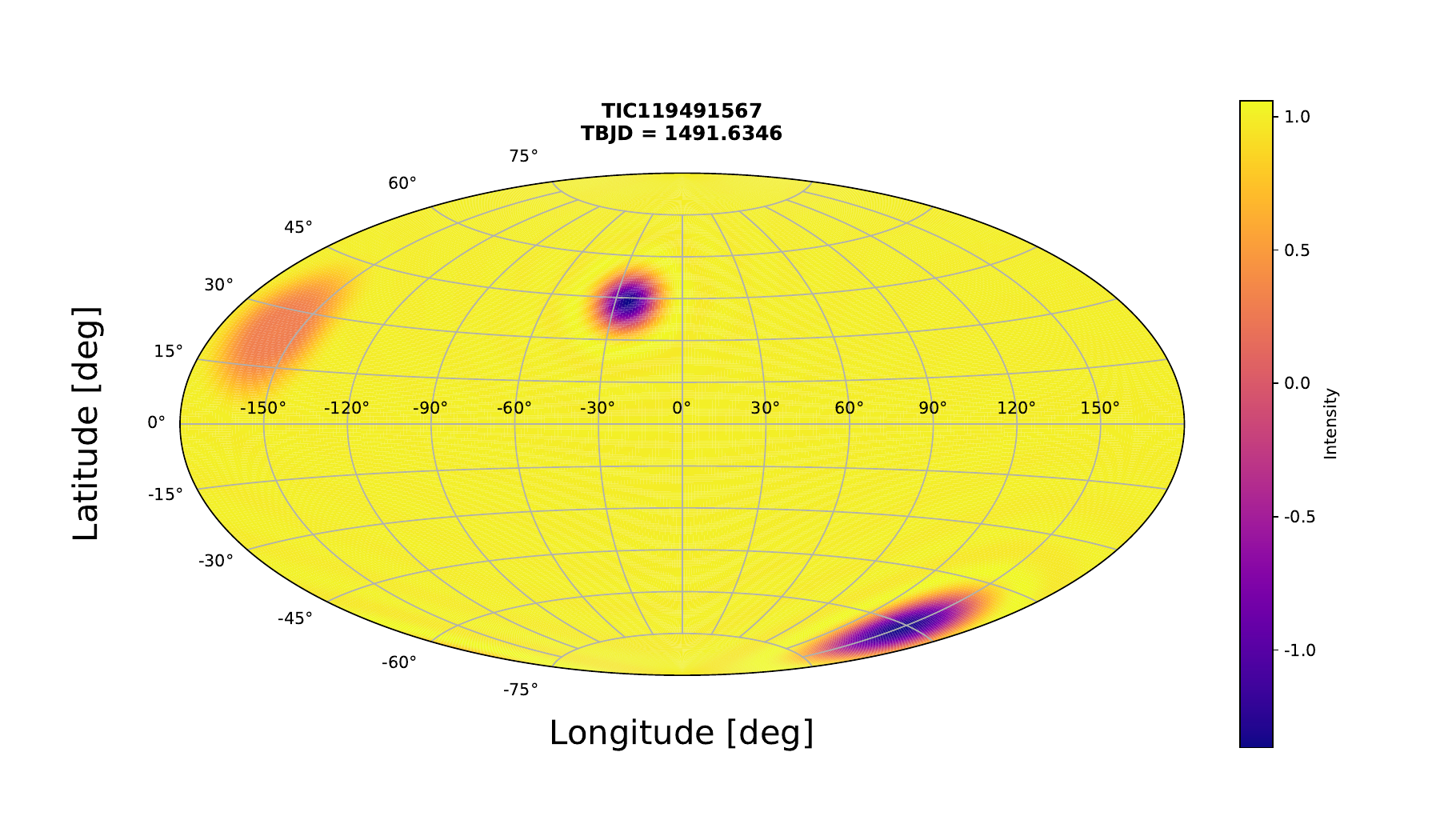}
\caption{ {Spot} 
 locations, sizes, and contrasts are illustrated using Aitoff projections for sectors 6 \linebreak  (\textbf{left panel}) and 7 (\textbf{right panel}). Dark regions indicate the presence of spots, and the strength of the spots has been scaled by the color map shown in the figures.}
\label{fig:fig4}
\end{figure}

The spot configurations, including locations, sizes, and temperature contrasts, are shown in Aitoff projections in Figure\,\ref{fig:fig4}, and the corresponding parameters are summarized in Table\,\ref{detailed-spot-table}. The latitudes of the spots remain almost unchanged, indicating that the spot locations on the stellar surface do not vary significantly. The longitudes differ, as expected, due to the rotation of the object and mapping them at different rotational phases. Even the longitudinal distance between the spots changes noticeably from one sector to another, suggesting differential surface rotation. However, Figure \ref{fig:fig4} shows that spots 1 and 2 shift slightly towards lower latitudes, while spot 3 shifts towards higher latitudes ($\lesssim$7$^{\circ}$). The contrast of spot 1 changes from sector 6 to sector 7, while the contrasts of spots 2 and 3 remain nearly constant, as indicated by the color bar. These results support the findings from our wavelet analysis. The $\sim$10-d periodicity and associated harmonics observed in the wavelet maps primarily originate from spots 1 and 2. In contrast, the polar location of spot 3 has little impact on the light curve morphology. Although the individual spot sizes vary between the two sectors, the total spot coverage does not change significantly (approximately $\sim$0.4\%). This signifies that spot modulation might be present. The value of $\chi^2$ $>$ 1 suggests the presence of small-scale variations with timescales shorter than the stellar rotation period, resulting in features in the light curve that the model light curves do not fully capture. Nevertheless, the high signal-to-noise ratio (SNR) values confirm that the modeled light curve is robust and reliable, which is consistent with the quality of the TESS light curves.

In the wavelet diagram, persistent horizontal bands correspond to stable periodic signals, indicative of coherent stellar rotation. Variations in dominant periodicities or light curve morphology over time may reflect evolving spot distributions, longitudinal migration of active regions, or differential rotation. In this case, the light curve morphology remains relatively unchanged across sectors, consistent with modest changes in spot configuration and coverage, particularly in longitudinal positions.

In this modeling, {\sc BASSMAN} assumes spots are spherical on the stellar surface. During the modeling of a single sector of TESS observations, spots are assumed to be static, i.e., they do not evolve in size, position, or contrast within that time frame. The only parameter that changes over time is the longitude, due to stellar rotation. Due to the degeneracy between spot contrast and size in single-band photometry, the model uses spot amplitude, i.e., the amount of flux blocked, rather than contrast. Spot amplitude is defined based on the range of flux variation; size is estimated based on the ingress and egress duration. Markov Chain Monte Carlo (MCMC) sampling is used to optimize the fit of the spot model to the observed light curve. So, small spot size with proper amplitude can generate significant variability in the light curve. Although this modeling approach does not convincingly establish that the small spots alone are responsible for the photometric modulation, and CP stars have been known to have larger spots, we have conducted a preliminary analysis of spot location on the stellar surface here. For example, if the contrast were lower for the spots, their size would increase, which could be the case for these CP stars. The `size' in our tables refers to the area needed to reproduce the observed light curve modulation under a reasonable assumed contrast derived from empirical relations. We emphasize that while the individual spots may be relatively small, the total spotted area on the visible hemisphere, combined with high contrast and favorable geometries (i.e., near-equatorial spots and favorable inclination), is sufficient to produce the observed light curve modulations. Our modeling also reproduces the phase and shape of the modulation across multiple TESS sectors, reinforcing the credibility of the fit. We acknowledge that, due to the degeneracy between spot size and contrast in single-color data, alternative models with slightly larger, lower-contrast spots cannot be fully ruled out. However, the solutions presented represent the best-fit results of a Bayesian sampling process (via MCMC) within a physically motivated parameter space, constrained by the theoretical values (\mbox{Equations (6) and (7)}), observed flux variations, and stellar parameters.

It is important to note that modeling stellar surface features is inherently degenerate~[\citealp{2013ApJS..205...17W}], as multiple spot configurations can produce light curves that are nearly indistinguishable. Therefore, any derived solution must be interpreted cautiously, with due consideration for uncertainties and degeneracies. While a single spot configuration may not fully capture the true surface geometry, consistent modeling across multiple light curves can reveal meaningful trends and correlations that reflect real physical processes on the stellar surface. Lastly, the BASSMAN code is typically employed for investigating surface spot distributions in late-type stars. In the present study, we extend its application to AL\,Col, a massive star. Given the fundamental differences in a spot's origin between late-type and massive stars, the results derived from this approach should be interpreted with due caution. 

\begin{table}[H]
\footnotesize
\caption{ {Comprehensive} 
 spot model parameters and associated diagnostics for  
AL\,Col.}
\label{detailed-spot-table}
\begin{tabularx}{\textwidth}{LCrrccCR}
\toprule
 \textbf{Sec. No.}& \textbf{Spot No.}& \textbf{Lat. (deg)} & \textbf{Long.  (deg)}& \textbf{Spot  Size (\%)}& \textbf{Analytical  Spot Size (\%)}& \boldmath{\textbf{$\chi^2$}} & \textbf{SNR} \\

\midrule
  & 1 & 49.09 & $-$120.05  & 1.49 & & & \\
 6 & 2 & 20.86 & 106.53  & 1.99 & 5.77 & 4.42 & 1657.84 \\
  & 3 & $-$65.32 & $-$85.27  & 1.00 & & & \\
  & 1 & 42.98 & $-$24.91   & 1.00 & & & \\
 7 & 2 & 23.92 & $-$157.87 & 2.49 & 5.68 & 5.89 & 1455.62 \\
  & 3 & $-$61.24 & 149.79  & 1.39 & & & \\
\bottomrule
\end{tabularx}
\end{table}

\vspace{-12pt}
\section{Discussion and Conclusions}
\label{Sec:conclusion}

This study presents a comprehensive photometric and high-resolution spectroscopic analysis of the Ap/Bp star AL\,Col (HD\,46462). Frequency analysis of the \tess\ light curve revealed a dominant peak at \( \nu = 0.09655\,\mathrm{d}^{-1} \), corresponding to a rotational period of \( P_{\mathrm{rot}} = 10.35733 \) d, in close agreement with the value of \( P_{\mathrm{rot}} = 10.363 \) d reported by \citet{2020MNRAS.493.3293B}. The phase-folded light curve exhibits a clear quasi-sinusoidal shape, characteristic of \(\alpha^2\)\,CVn-type variability commonly seen in Ap/Bp stars. Assuming rigid-body rotation and adopting a spectroscopically derived \vsini\ of 16\,\kms, we estimate an inclination angle of the stellar rotation axis of $i \approx 60$$^{\circ}$.

A notable discrepancy was noticed between the effective temperature derived from SED fitting (\( T_{\mathrm{eff}} = 11,750\,\mathrm{K} \)) and that obtained from high-resolution spectroscopic analysis using the SME tool (\( T_{\mathrm{eff}} = 13,814 \pm 400\,\mathrm{K} \)). This is common in CP stars because of flux redistribution in the presence of chemical spots. Flux are being absorbed (maximum in ultraviolet) by chemical elements present in the photosphere and getting emitted in a longer wavelength (generally in IR), which changes the shape of the SED \citep{1991A&A...242..199L}. This highlights the importance of detailed spectral synthesis for accurate atmospheric parameter determination in CP stars. The core of the strong lines like the Balmer regions and the Fe line in the Mg triplet region are not properly fitted. NLTE analysis is one of the possible solutions for fitting the core region of Balmer lines. Also, in the H$\alpha$ region, the presence of telluric lines in the wings occasionally leads to poor agreement between the observed and synthetic spectra.  {However,} 
 for metal lines (e.g., Fe II in Figure \ref{fig:fig6}), the line profile of the core region varies (spikes at the center) due to chemical spots. To model these line profile variations, time-series spectroscopy is essential.

Using synthetic spectrum fitting, we derived abundances for 25 chemical elements across various ionization stages. Our results show that O\,\textsc{ii}, Mg\,\textsc{ii}, S\,\textsc{ii}, and Ca\,\textsc{ii} have abundances close to solar, while He\,\textsc{i} is significantly underabundant by about 0.5\,dex, possibly due to diffusion processes \citep{Bailey_2014}. Rare earth elements (REEs), on the other hand, exhibit substantial overabundances up to 6\,dex, consistent with the typical chemical peculiarities observed in Ap/Bp stars.

At \teff\,$ = 13,814 \pm 400$\,K and \logg\,$=4.09\pm0.08$\,dex, the atmosphere lies in a regime where departures from the LTE are expected.  Nevertheless, all abundances reported here were derived strictly under LTE, and the formal uncertainties quoted in Table \ref{tab:table_abundances} do not include any allowance for NLTE offsets.  Literature grids predict downward corrections of $\approx$0.10--0.15\,dex for He\,{\sc i} triplet lines and up to $-$0.20\,dex for C\,{\sc ii}/N\,{\sc ii}/O\,{\sc ii} lines \citep{Przybilla2005, NievaPrzybilla2012}.  Mg\,{\sc ii} and Si\,{\sc ii/iii} resonance lines are expected to shift upward by +0.05--0.10\,dex, whereas Fe\,{\sc ii} and Cr\,{\sc ii} remain within $|\Delta\log\epsilon|\lesssim0.05$\,dex thanks to efficient collisional \mbox{coupling~\citep{Lind2012, Mashonkina2016, Sitnova2020}}.  For the observed rare-earth ions, Y\,{\sc ii}, Ce\,{\sc ii/iii}, Pr\,{\sc ii/iii}, Nd\,{\sc ii/iii}, and Eu\,{\sc iii}, published NLTE studies suggest mainly negative corrections of 0.05--0.15\,dex \citep{Mashonkina2005, Ryabchikova2015}.  These potential systematic offsets will be explored in a future re-analysis employing full NLTE spectrum synthesis; until then, the LTE abundances and their errors should be regarded as lower-limit uncertainties.

Although magnetic measurements were not obtained in the present study, it is well established that magnetic fields in Ap/Bp stars suppress convective and turbulent motions, enabling the radiative accumulation of heavy elements like REEs in the photosphere. Future spectropolarimetric observations are essential to constrain the magnetic field geometry and understand its role in shaping the observed abundance anomalies.

A significant discrepancy was also noted between the abundances derived from Si\,\textsc{ii} and Si\,\textsc{iii} lines, consistent with the expected vertical stratification of chemical elements in magnetic atmospheres. In particular, Si\,\textsc{iii} lines yield a considerably higher abundance than Si\,\textsc{ii}, a common feature in magnetic chemically peculiar stars (e.g., \citep{refId0}). This underscores the need for future non-LTE analyses and stratified atmospheric modeling to better understand the vertical abundance gradients.

By locating our target in the H--R diagram, we estimate the stellar mass to be\linebreak   \( M = 4.0 \pm 0.4\,{\rm M}_\odot \) and the radius to be \( R = 3.74 \pm 0.48\,{\rm R}_\odot \), indicating that the star has evolved slightly beyond the main sequence. The isochrone fitting suggests an age of \( 0.12 \pm 0.01 \)\,Gyr.

To probe surface activity, we modeled starspots using \tess\ time-series photometry from sectors 6 and 7 via the BASSMAN code. The best-fit three-spot model revealed the presence of cool surface features at different longitudes and latitudes. Temporal evolution in spot sizes and contrasts, most notably an increase in the strength of one spot, suggests the presence of differential rotation and evolving surface magnetic activity. These findings are tentative though corroborated with our wavelet analysis, which detected persistent periodicities consistent with rotational modulation by surface spots.

Collectively, these results provide a coherent picture of AL\,Col as a moderately evolved magnetic Ap/Bp star exhibiting strong chemical peculiarities, rotational modulation due to starspots, and potential surface differential rotation. Follow-up spectropolarimetric and non-LTE spectroscopic analyses will be instrumental in refining our understanding of its magnetic topology and atmospheric structure.

\vspace{6pt}

\authorcontributions{ { } 
}

\funding{ {The} 
 authors acknowledge the financial support received from the BRICS grant under the project DST/ICD/BRICS/Call-5/SAPTARISI/2023 (G). J.A.R. acknowledges support by Vicerrector\'{i}a de Investigaci\'{o}n y Desarrollo (VRID) Postdoctoral DICA N181/23. OT acknowledges financial support from the SEISMIC project, the Max Planck--Humboldt Research Unit in collaboration with the Max Planck Institutes for Astrophysics (MPA) and Solar System Research (MPS), and Kyambogo University, as well as from the International Science Programme (ISP) at Uppsala University. AD acknowledges the financial support received from the DST-INSPIRE Fellowship Programme (DST/INSPIRE Fellowship/2020/IF200245). KT acknowledges support by the National Natural Science Foundation of China (NSFC) under grant no. 12261141689.}

\dataavailability{The data presented in this study are available on
request from the corresponding author.}

\acknowledgments{
 {This} 
 study uses data from the TESS missions that are accessible at the Mikulski Archive for Space Telescopes (MAST) operated by NASA. The wavelet maps and autocorrelation function were created using the code provided by  {Rafel Garcia} 
 and Savita Mathur. The spectroscopic analysis for this study is based on the data available in the European Southern Observatory (ESO) Science Archive Facility under ESO Prog ID 080.C-0032(A).}

\conflictsofinterest{The authors declare no conflicts of interest.}

\begin{adjustwidth}{-\extralength}{0cm}
\printendnotes[custom] 

\reftitle{References}


\begin{thebibliography}{999}

\bibitem[{Lueftinger} et~al.(2003){Lueftinger}, {Wade}, and
  {Weiss}]{2003EAS.....9..249L}
{Lueftinger}, T.; {Wade}, G.A.; {Weiss}, W.W.
\newblock {Magnetic Fields in B, (peculiar) A, and F stars recent results from
  MuSiCoS Data}.
\newblock In \emph{Proceedings of the EAS Publications Series}; {Arnaud}, J.,
  {Meunier}, N., Eds.; EAS Publications Series;  {Cambridge University Press:  Cambridge, UK, 2003;} 
 p. 249, Volume 9.

\bibitem[{Lehmann} et~al.(2007){Lehmann}, {Tkachenko}, {Fraga}, {Tsymbal}, and
  {Mkrtichian}]{2007A&A...471..941L}
{Lehmann}, H.; {Tkachenko}, A.; {Fraga}, L.; {Tsymbal}, V.; {Mkrtichian}, D.E.
\newblock {The helium weak silicon star HR  {7224.} 
 II. Doppler
  Imaging analysis}. \emph{Astron. Astrophys.} 
\newblock {\bf 2007}, \emph{471},~941--949.
\newblock {\url{https://doi.org/10.1051/0004-6361:20077700}}.

\bibitem[{Preston}(1974)]{1974ARA&A..12..257P}
{Preston}, G.W.
\newblock {The chemically peculiar stars of the upper main sequence.} \emph{Annu. Rev. Astron. Astrophys.}
\newblock  {\bf 1974}, {\emph{12}},~257--277.
\newblock {\url{https://doi.org/10.1146/annurev.aa.12.090174.001353}}.

\bibitem[{Maitzen}(1984)]{1984A&A...138..493M}
{Maitzen}, H.M.
\newblock {A note on the relation between magnetic fields and the lambda 5200
  feature in helium weak-stars.} \emph{Astron. Astrophys.}
\newblock {\bf 1984}, {\em 138},~493--494.

\bibitem[{Paunzen}(2004)]{2004IAUS..224..443P}
{Paunzen}, E.
\newblock {The {\ensuremath{\lambda}} Bootis stars}.
\newblock In \emph{Proceedings of the A-Star Puzzle}; {Zverko}, J., {Ziznovsky},
  J., {Adelman}, S.J., {Weiss}, W.W., Eds.; 
 {Cambridge University Press: Cambridge, UK,} 
 2004; Volume 224,  {{\em IAU
  Symposium},} 
 pp. 443--450.
\newblock {\url{https://doi.org/10.1017/S1743921304004867}}.

\bibitem[{Ghazaryan} et~al.(2018){Ghazaryan}, {Alecian}, and
  {Hakobyan}]{2018CoBAO..65..223G}
{Ghazaryan}, S.; {Alecian}, G.; {Hakobyan}, A.A.
\newblock {Statistical Analysis of the New Catalogue of CP Stars}.
\newblock {\em Commun. Byurakan Astrophys. Obs.} {\bf
  2018}, {\em 65},~223-- {227.} 
\newblock {\url{https://doi.org/10.52526/25792776-2018.2.2-223}}.

\bibitem[{Auri{\`e}re} et~al.(2007){Auri{\`e}re}, {Wade}, {Silvester},
  {Ligni{\`e}res}, {Bagnulo}, {Bale}, {Dintrans}, {Donati}, {Folsom},
  {Gruberbauer}, {Hui Bon Hoa}, {Jeffers}, {Johnson}, {Landstreet},
  {L{\`e}bre}, {Lueftinger}, {Marsden}, {Mouillet}, {Naseri}, {Paletou},
  {Petit}, {Power}, {Rincon}, {Strasser}, and {Toqu{\'e}}]{2007A&A...475.1053A}
{Auri{\`e}re}, M.; {Wade}, G.A.; {Silvester}, J.; {Ligni{\`e}res}, F.;
  {Bagnulo}, S.; {Bale}, K.; {Dintrans}, B.; {Donati}, J.F.; {Folsom}, C.P.;
  {Gruberbauer}, M.;  et~al.
\newblock {Weak magnetic fields in Ap/Bp stars. Evidence for a dipole field
  lower limit and a tentative interpretation of the magnetic dichotomy}. \emph{Astron. Astrophys.}
\newblock{\bf 2007}, {\em 475},~1053-- {1065.} 
\newblock {\url{https://doi.org/10.1051/0004-6361:20078189}}.

\bibitem[{Kochukhov}(2011)]{2011IAUS..273..249K}
{Kochukhov}, O.
\newblock {The spots on Ap stars}.
\newblock In \emph{Proceedings of the Physics of Sun and Star Spots}; {Prasad
  Choudhary}, D., {Strassmeier}, K.G., Eds.; 
   {Cambridge University Press: Cambridge, UK,} 
   2011, Volume 273,  {{\em IAU
  Symposium},} 
   pp. 249-- {255.} 
\newblock {\url{https://doi.org/10.1017/S1743921311015328}}.

\bibitem[Shultz et~al.(2019)Shultz, Wade, Alecian, et~al.]{2019MNRAS.490..274S}
Shultz, M.; Wade, G.A.; Alecian, E.;   {et~al.} 
\newblock The incidence and properties of magnetic fields in early B-type
  stars.
\newblock {\em   {Mon. Not. R. Astron. Soc.} 
} {\bf 2019}, {\em 490},~274-- {296.} 
\newblock {\url{https://doi.org/10.1093/mnras/stz2546}}.

\bibitem[Braithwaite and Spruit(2014)]{2014IAUS..302..338B}
Braithwaite, J.; Spruit, H.C.
\newblock Fossil magnetic fields in stars: Origin and evolution.
\newblock {\em Proc. Int. Astron. Union} {\bf 2014},
  {\em 302},~338--347.
\newblock {\url{https://doi.org/10.1017/S1743921314002427}}.

\bibitem[{Michaud}(1970)]{1970ApJ...160..641M}
{Michaud}, G.
\newblock {Diffusion Processes in Peculiar a Stars}. \emph{Astrophys. J.}
\newblock  {\bf 1970}, {\em 160},~641.
\newblock {\url{https://doi.org/10.1086/150459}}.

\bibitem[Richer et~al.(2005)Richer, Michaud, and Turcotte]{2005ApJ...625..548R}
Richer, J.; Michaud, G.; Turcotte, S.
\newblock Implications of Diffusion for the Surface Composition of A and B
  Stars.
\newblock {\em ApJ} {\bf 2005}, {\em 625},~548--556.
\newblock {\url{https://doi.org/10.1086/429648}}.

\bibitem[{Michaud} et~al.(2015){Michaud}, {Alecian}, and
  {Richer}]{2015ads..book.....M}
{Michaud}, G.; {Alecian}, G.; {Richer}, J.
\newblock {\em {Atomic Diffusion in Stars}};  {Springer: Berlin/Heidelberg, Germany,} 
 2015.
\newblock {\url{https://doi.org/10.1007/978-3-319-19854-5}}.

\bibitem[Alecian et~al.(2020)]{2019EAS....82..345A}
{Alecian}, E. and {Villebrun}, F. and {Grunhut}, J. and {Hussain}, G. and {Neiner}, C. and {Wade}, G.~A. 
\newblock Magnetism in intermediate-mass stars: From fossil fields to dynamo
  action.
\newblock {\em {EAS Publications Series} 
} {\bf 2019}, {\em 82},~345--355. 
\newblock {\url{https://ui.adsabs.harvard.edu/abs/2019EAS....82..345A}}.

\bibitem[{Babcock}(1949)]{1949Obs....69..191B}
{Babcock}, H.W.
\newblock {Stellar magnetic fields and rotation}.
\newblock {\em Observatory} {\bf 1949}, {\em 69},~191--192.

\bibitem[{Stibbs}(1950)]{1950Natur.165..195S}
{Stibbs}, D.W.N.
\newblock {A Light Curve for the Magnetic Variable Star HD125248}. \emph{Nature}
\newblock {\bf 1950}, {\em 165},~195.
\newblock {\url{https://doi.org/10.1038/165195a0}}.

\bibitem[{Shulyak} et~al.(2010){Shulyak}, {Krti{\v{c}}ka},
  {Mikul{\'a}{\v{s}}ek}, {Kochukhov}, and {L{\"u}ftinger}]{2010A&A...524A..66S}
{Shulyak}, D.; {Krti{\v{c}}ka}, J.; {Mikul{\'a}{\v{s}}ek}, Z.; {Kochukhov}, O.;
  {L{\"u}ftinger}, T.
\newblock {Modelling the light variability of the Ap star $\epsilon$ Ursae
  Majoris}. \emph{Astron. Astrophys.}
\newblock {\bf 2010}, {\em 524},~ {A66.} 
\newblock {\url{https://doi.org/10.1051/0004-6361/201015094}}.

\bibitem[Kochukhov et~al.(2012)Kochukhov, Wade, and
  Shulyak]{2012A&A...542A.116K}
Kochukhov, O.; Wade, G.A.; Shulyak, D.
\newblock Zeeman Doppler imaging of magnetic Ap stars: A practical approach.
\newblock {\em Astron. Astrophys.} {\bf 2012}, {\em 542},~A116.
\newblock {\url{https://doi.org/10.1051/0004-6361/201118758}}.

\bibitem[{Kurtz} et~al.(2006){Kurtz}, {Elkin}, {Cunha}, {Mathys}, {Hubrig},
  {Wolff}, and {Savanov}]{2006Kurtz}
{Kurtz}, D.W.; {Elkin}, V.G.; {Cunha}, M.S.; {Mathys}, G.; {Hubrig}, S.;
  {Wolff}, B.; {Savanov}, I.
\newblock {The discovery of 8.0-min radial velocity variations in the strongly
  magnetic cool Ap star HD154708, a new roAp star}.
\newblock {\em  {Mon. Not. R. Astron. Soc.} 
} {\bf 2006}, {\em 372},~286--292.
\newblock {\url{https://doi.org/10.1111/j.1365-2966.2006.10857.x}}.

\bibitem[Saio et~al.(2012)Saio, Kurtz, Takata, et~al.]{2012MNRAS.420..283S}
{Saio}, H. and {Gruberbauer}, M. and {Weiss}, W.~W. and {Matthews}, J.~M. and {Ryabchikova}, T. 
\newblock Pulsation models for the roAp star HD 134214
\newblock {\em{Mon. Not. R. Astron. Soc.} 
} {\bf 2012}, {\em 420},~283--290 
\newblock {\url{https://ui.adsabs.harvard.edu/abs/2012MNRAS.420..283S}}.

\bibitem[{Joshi} and {Joshi}(2015)]{2015JApA...36...33J}
{Joshi}, S.; {Joshi}, Y.C.
\newblock {Asteroseismology of Pulsating Stars}.
\newblock {\em J. Astrophys. Astron.} {\bf 2015}, {\em
  36},~33-- {80.} 
\newblock {\url{https://doi.org/10.1007/s12036-015-9327-z}}.

\bibitem[{Holdsworth} et~al.(2021){Holdsworth}, {Cunha}, {Kurtz}, {Antoci},
  {Hey}, {Bowman}, {Kobzar}, {Buzasi}, {Kochukhov}, {Niemczura}, {Ozuyar},
  {Shi}, {Szab{\'o}}, {Samadi-Ghadim}, {Bogn{\'a}r}, {Fox-Machado}, {Khalack},
  {Lares-Martiz}, {Lovekin}, {Miko{\l}ajczyk}, {Mkrtichian}, {Pascual-Granado},
  {Paunzen}, {Richey-Yowell}, {S{\'o}dor}, {Sikora}, {Yang}, {Brunsden},
  {David-Uraz}, {Derekas}, {Garc{\'\i}a Hern{\'a}ndez}, {Guzik}, {Hatamkhani},
  {Handberg}, {Lambert}, {Lampens}, {Murphy}, {Monier}, {Pollard},
  {Quitral-Manosalva}, {Ram{\'o}n-Ballesta}, {Smalley}, {Stateva}, and
  {Vanderspek}]{2021MNRAS.506.1073H}
{Holdsworth}, D.L.; {Cunha}, M.S.; {Kurtz}, D.W.; {Antoci}, V.; {Hey}, D.R.;
  {Bowman}, D.M.; {Kobzar}, O.; {Buzasi}, D.L.; {Kochukhov}, O.; {Niemczura},
  E.;  et~al.
\newblock {TESS cycle 1 observations of roAp stars with 2-min cadence data}. \emph{Mon. Not. R. Astron. Soc.}
\newblock  {\bf 2021}, {\em 506},~1073-- {1110.} 
\newblock {\url{https://doi.org/10.1093/mnras/stab1578}}.

\bibitem[{Kurtz}(2022)]{2022ARA&A..60...31K}
{Kurtz}, D.W.
\newblock {Asteroseismology Across the Hertzsprung-Russell Diagram}. \emph{Annu. Rev. Astron. Astrophys.}
\newblock  {\bf 2022}, {\em 60},~31--71.
\newblock {\url{https://doi.org/10.1146/annurev-astro-052920-094232}}.

\bibitem[{Martinez} et~al.(1991){Martinez}, {Kurtz}, and
  {Kauffmann}]{1991MNRAS.250..666M}
{Martinez}, P.; {Kurtz}, D.W.; {Kauffmann}, G.M.
\newblock {The Cape rapidly oscillating star survey-I. First results.} \emph{Mon. Not. R. Astron. Soc.}
\newblock  {\bf 1991}, {\em 250},~666.
\newblock {\url{https://doi.org/10.1093/mnras/250.4.666}}.

\bibitem[{Martinez} et~al.(2001){Martinez}, {Kurtz}, {Ashoka}, {Chaubey},
  {Girish}, {Gupta}, {Joshi}, {Kasturirangan}, {Sagar}, and
  {Seetha}]{2001AA...371.1048M}
{Martinez}, P.; {Kurtz}, D.W.; {Ashoka}, B.N.; {Chaubey}, U.S.; {Girish}, V.;
  {Gupta}, S.K.; {Joshi}, S.; {Kasturirangan}, K.; {Sagar}, R.; {Seetha}, S.
\newblock {The Naini Tal - Cape survey for pulsations in chemically peculiar
  A-type stars. I. Methods and preliminary results}. \emph{Astron. Astrophys.}
\newblock{\bf 2001}, {\em 371},~1048--1055.
\newblock {\url{https://doi.org/10.1051/0004-6361:20010432}}.

\bibitem[{Joshi} et~al.(2003){Joshi}, {Girish}, {Sagar}, {Kurtz}, {Martinez},
  {Kumar}, {Seetha}, {Ashoka}, and {Zhou}]{2003MNRAS.344..431J}
{Joshi}, S.; {Girish}, V.; {Sagar}, R.; {Kurtz}, D.W.; {Martinez}, P.; {Kumar},
  B.; {Seetha}, S.; {Ashoka}, B.N.; {Zhou}, A.
\newblock {Discovery of unusual pulsations in the cool, evolved Am stars HD
  98851 and HD 102480}. \emph{Mon. Not. R. Astron. Soc.}
\newblock  {\bf 2003}, {\em 344},~431-- {438.} 
\newblock {\url{https://doi.org/10.1046/j.1365-8711.2003.06823.x}}.

\bibitem[{Joshi} et~al.(2006){Joshi}, {Mary}, {Martinez}, {Kurtz}, {Girish},
  {Seetha}, {Sagar}, and {Ashoka}]{2006A&A...455..303J}
{Joshi}, S.; {Mary}, D.L.; {Martinez}, P.; {Kurtz}, D.W.; {Girish}, V.;
  {Seetha}, S.; {Sagar}, R.; {Ashoka}, B.N.
\newblock {The Nainital-Cape Survey. II. Report for pulsation in five
  chemically peculiar A-type stars and presentation of 140 null results}. \emph{Astron. Astrophys.}
\newblock  {\bf 2006}, {\em 455},~303-- {313.} 
\newblock {\url{https://doi.org/10.1051/0004-6361:20064970}}.

\bibitem[{Joshi} et~al.(2009){Joshi}, {Mary}, {Chakradhari}, {Tiwari}, and
  {Billaud}]{2009A&A...507.1763J}
{Joshi}, S.; {Mary}, D.L.; {Chakradhari}, N.K.; {Tiwari}, S.K.; {Billaud}, C.
\newblock {The Nainital-Cape Survey. III. A search for pulsational variability
  in chemically peculiar stars}. \emph{Astron. Astrophys.}
\newblock {\bf 2009}, {\em 507},~1763-- {1784.} 
\newblock {\url{https://doi.org/10.1051/0004-6361/200912382}}.

\bibitem[{Joshi} et~al.(2010){Joshi}, {Ryabchikova}, {Kochukhov}, {Sachkov},
  {Tiwari}, {Chakradhari}, and {Piskunov}]{2010MNRAS.401.1299J}
{Joshi}, S.; {Ryabchikova}, T.; {Kochukhov}, O.; {Sachkov}, M.; {Tiwari}, S.K.;
  {Chakradhari}, N.K.; {Piskunov}, N.
\newblock {Time-resolved photometric and spectroscopic analysis of the luminous
  Ap star HD103498}. \emph{Mon. Not. R. Astron. Soc.}
\newblock  {\bf 2010}, {\em 401},~1299-- {1307.} 
\newblock {\url{https://doi.org/10.1111/j.1365-2966.2009.15725.x}}.

\bibitem[{Joshi} et~al.(2012){Joshi}, {Semenko}, {Martinez}, {Sachkov},
  {Joshi}, {Seetha}, {Chakradhari}, {Mary}, {Girish}, and
  {Ashoka}]{2012MNRAS.424.2002J}
{Joshi}, S.; {Semenko}, E.; {Martinez}, P.; {Sachkov}, M.; {Joshi}, Y.C.;
  {Seetha}, S.; {Chakradhari}, N.K.; {Mary}, D.L.; {Girish}, V.; {Ashoka}, B.N.
\newblock {A spectroscopic analysis of the chemically peculiar star HD 207561}. \emph{Mon. Not. R. Astron. Soc.}
\newblock  {\bf 2012}, {\em 424},~2002-- {2008.} 
\newblock {\url{https://doi.org/10.1111/j.1365-2966.2012.21340.x}}.

\bibitem[{Joshi} et~al.(2016){Joshi}, {Martinez}, {Chowdhury}, {Chakradhari},
  {Joshi}, {van Heerden}, {Medupe}, {Kumar}, and {Kuhn}]{2016A&A...590A.116J}
{Joshi}, S.; {Martinez}, P.; {Chowdhury}, S.; {Chakradhari}, N.K.; {Joshi},
  Y.C.; {van Heerden}, P.; {Medupe}, T.; {Kumar}, Y.B.; {Kuhn}, R.B.
\newblock {The Nainital-Cape Survey. IV. A search for pulsational variability
  in 108 chemically peculiar stars}. \emph{Astron. Astrophys.}
\newblock {\bf 2016}, {\em 590},~ {A116.} 
\newblock {\url{https://doi.org/10.1051/0004-6361/201527242}}.

\bibitem[{Joshi} et~al.(2017){Joshi}, {Semenko}, {Moiseeva}, {Sharma}, {Joshi},
  {Sachkov}, {Singh}, and {Yerra}]{2017MNRAS.467..633J}
{Joshi}, S.; {Semenko}, E.; {Moiseeva}, A.; {Sharma}, K.; {Joshi}, Y.C.;
  {Sachkov}, M.; {Singh}, H.P.; {Yerra}, B.K.
\newblock {High-resolution Spectroscopy and Spectropolarimetry of Selected
  {\ensuremath{\delta}}-Sct Pulsating Variables}. \emph{Mon. Not. R. Astron. Soc.}
\newblock {\bf 2017}, {\em 467},~633-- {645.} 
\newblock {\url{https://doi.org/10.1093/mnras/stx087}}.

\bibitem[{Joshi} et~al.(2022){Joshi}, {Trust}, {Semenko}, {Williams},
  {Lampens}, {De Cat}, {Vermeylen}, {Holdsworth}, {Garc{\'\i}a}, {Mathur},
  {Santos}, {Mkrtichian}, {Goswami}, {Cuntz}, {Yadav}, {Sarkar}, {Bhatt},
  {Kahraman Ali{\c{c}}avu{\c{s}}}, {Nhlapo}, {Lund}, {Goswami}, {Savanov},
  {Jorissen}, {Jurua}, {Avvakumova}, {Dmitrienko}, {Chakradhari}, {Das},
  {Chowdhury}, {Abedigamba}, {Yakunin}, {Letarte}, and
  {Karinkuzhi}]{2022MNRAS.510.5854J}
{Joshi}, S.; {Trust}, O.; {Semenko}, E.; {Williams}, P.E.; {Lampens}, P.; {De
  Cat}, P.; {Vermeylen}, L.; {Holdsworth}, D.L.; {Garc{\'\i}a}, R.A.; {Mathur},
  S.;  et~al.
\newblock {Study of chemically peculiar stars--I. High-resolution spectroscopy
  and K2 photometry of Am stars in the region of M44}. \emph{Mon. Not. R. Astron. Soc.}
\newblock  {\bf 2022}, {\em 510},~ {5854.} 
\newblock {\url{https://doi.org/10.1093/mnras/stab3158}}.

\bibitem[Buysschaert et~al.(2019)Buysschaert, Neiner, De~Cat,
  et~al.]{2019A&A...622A..67B}
{Buysschaert}, B. and {Neiner}, C. and {Martin}, A.~J. and {Oksala}, M.~E. and {Aerts}, C. and {Tkachenko}, A. and {Alecian}, E. and {MiMeS Collaboration} 
\newblock Magnetic characterization and variability study of the magnetic SPB star o Lupi \emph{Astron. Astrophys.}
\newblock {\bf 2019}, {\em 622},{A67.} 
\newblock {\url{https://ui.adsabs.harvard.edu/abs/2019A&A...622A..67B}}.

\bibitem[Holdsworth et~al.(2021)Holdsworth, Smalley, Kurtz,
  et~al.]{10.1093/mnras/stad3800}
Holdsworth, D L and Cunha, M S and Lares-Martiz, M and Kurtz, D W and Antoci, V and Barceló Forteza, S and De Cat, P and Derekas, A and Kayhan, C and Ozuyar, D 
\newblock TESS Cycle 2 observations of roAp stars with 2-min cadence data. \emph{Mon. Not. R. Astron. Soc.}
\newblock {\bf 2023}, {\em 527},~9548--{9580.} 
\newblock {\url{https://doi.org/10.1093/mnras/stad3800}}.

\bibitem[Bagnulo et~al.(2023)Bagnulo, Landstreet, Power,
  et~al.]{2023A&A...671A.154B}
Bagnulo, S.; Landstreet, J.D.; Power, J.;  {et~al.} 
\newblock Magnetic chemically peculiar stars in the TESS era: Rotation and
  magnetic field from photometry and spectroscopy. \emph{Astron. Astrophys.}
\newblock {\bf 2023}, {\em 671},~A154.
\newblock {\url{https://doi.org/10.1051/0004-6361/202245820}}.

\bibitem[Wade et~al.(2014)Wade, Alecian, Grunhut, et~al.]{2014IAUS..302..265W}
Wade, G.A.; Alecian, E.; Grunhut, J.H.;   {et~al.} 
\newblock The MiMeS survey of magnetism in massive stars: Introduction and
  overview.
\newblock {\em Proc. Int. Astron. Union} {\bf 2014},
  {\em 302},~265--270.
\newblock {\url{https://doi.org/10.1017/S1743921314002440}}.

\bibitem[Sikora et~al.(2020)Sikora, Wade, and Power]{2020MNRAS.499.5049S}
Sikora, J.; Wade, G.A.; Power, J.
\newblock Magnetism in Ap/Bp stars with rotational periods longer than 100 d.
\newblock {\em Mon. Not. R. Astron. Soc.} {\bf 2020},
  {\em 499},~5049--5068.
\newblock {\url{https://doi.org/10.1093/mnras/staa3149}}.

\bibitem[{Bernhard} et~al.(2020){Bernhard}, {H{\"u}mmerich}, and
  {Paunzen}]{2020MNRAS.493.3293B}
{Bernhard}, K.; {H{\"u}mmerich}, S.; {Paunzen}, E.
\newblock {New and improved rotational periods of magnetic CP stars from
  ASAS-3, KELT, and MASCARA data}. \emph{Mon. Not. R. Astron. Soc.}
\newblock {\bf 2020}, {\em 493},~3293-- {3330.} 
\newblock {\url{https://doi.org/10.1093/mnras/staa462}}.

\bibitem[Collaboration et~al.(2018)Collaboration, Cardoso, Hedges,
  Gully-Santiago, Saunders, Cody, Barclay, Hall, Sagear, Turtelboom,
  et~al.]{collaboration2018lightkurve}
Lightkurve Collaboration; Cardoso, J.; Hedges, C.; Gully-Santiago, M.; Saunders, N.;
  Cody, A.; Barclay, T.; Hall, O.; Sagear, S.; Turtelboom, E.;  et~al.
\newblock \emph{Lightkurve: Kepler and TESS Time Series Analysis in Python};
\newblock { {Astrophysics Source Code Library (ascl: 1812.013):} 
 } {2018}.

\bibitem[{Lenz} and {Breger}(2005)]{2005CoAst.146...53L}
{Lenz}, P.; {Breger}, M.
\newblock {Period04 User Guide}.
\newblock {\em CoAst} {\bf 2005}, {\em 146},~53--136.
\newblock {\url{https://doi.org/10.1553/cia146s53}}.

\bibitem[{Baran} and {Koen}(2021)]{2021AcA....71..113B}
{Baran}, A.S.; {Koen}, C.
\newblock {A Detection Threshold in the Amplitude Spectra Calculated from TESS
  Time-Series Data}. \emph{Acta Astron.}
\newblock {\bf 2021}, {\em 71},~113-- {121.} 
\newblock {\url{https://doi.org/10.32023/0001-5237/71.2.3}}.

\bibitem[{Bravo} et~al.(2014){Bravo}, {Roque}, {Estrela}, {Le{\~a}o}, and {De
  Medeiros}]{2014A&A...568A..34B}
{Bravo}, J.P.; {Roque}, S.; {Estrela}, R.; {Le{\~a}o}, I.C.; {De Medeiros},
  J.R.
\newblock {Wavelets: A powerful tool for studying rotation, activity, and
  pulsation in Kepler and CoRoT stellar light curves}. \emph{Astron. Astrophys.}
\newblock {\bf 2014}, {\em 568},~ {A34.} 
\newblock {\url{https://doi.org/10.1051/0004-6361/201323032}}.

\bibitem[{Mathur, S.} et~al.(2014){Mathur, S.}, {García, R. A.}, {Ballot, J.},
  {Ceillier, T.}, {Salabert, D.}, {Metcalfe, T. S.}, {Régulo, C.}, {Jiménez,
  A.}, and {Bloemen, S.}]{Mathur}
{Mathur, S.}; {García, R. A.}; {Ballot, J.}; {Ceillier, T.}; {Salabert,
  D.}; {Metcalfe, T. S.}; {Régulo, C.}; {Jiménez, A.}; {Bloemen, S.}
\newblock Magnetic activity of F stars observed by Kepler. \emph{Astron. Astrophys.}
\newblock {\bf 2014}, {\em 562},~A124.
\newblock {\url{https://doi.org/10.1051/0004-6361/201322707}}.

\bibitem[{Ochsenbein} et~al.(2000){Ochsenbein}, {Bauer}, and
  {Marcout}]{2000A&AS..143...23O}
{Ochsenbein}, F.; {Bauer}, P.; {Marcout}, J.
\newblock {The VizieR database of astronomical catalogues}. \emph{Astron. Astrophys. Suppl. Ser.}
\newblock  {\bf 2000}, {\em 143},~23-- {32.} 
\newblock {\url{https://doi.org/10.1051/aas:2000169}}.

\bibitem[{Paunzen}(2015)]{2015A&A...580A..23P}
{Paunzen}, E.
\newblock {A new catalogue of Str{\"o}mgren-Crawford uvby{$\beta$} photometry}. \emph{Astron. Astrophys.}
\newblock  {\bf 2015}, {\em 580},~ {A23.} 
\newblock {\url{https://doi.org/10.1051/0004-6361/201526413}}.

\bibitem[{H{\o}g} et~al.(2000){H{\o}g}, {Fabricius}, {Makarov}, {Bastian},
  {Schwekendiek}, {Wicenec}, {Urban}, {Corbin}, and
  {Wycoff}]{2000A&A...357..367H}
{H{\o}g}, E.; {Fabricius}, C.; {Makarov}, V.V.; {Bastian}, U.; {Schwekendiek},
  P.; {Wicenec}, A.; {Urban}, S.; {Corbin}, T.; {Wycoff}, G.
\newblock {Construction and verification of the Tycho-2 Catalogue}. \emph{Astron. Astrophys.}
\newblock  {\bf 2000}, {\em 357},~367--386.

\bibitem[{Gaia Collaboration} et~al.(2022){Gaia Collaboration}, {Montegriffo},
  {Bellazzini}, {De Angeli}, {Andrae}, {Barstow}, {Bossini}, {Bragaglia},
  {Burgess}, {Cacciari}, {Carrasco}, {Chornay}, {Delchambre}, {Evans},
  {Fouesneau}, {Fremat}, {Garabato}, {Jordi}, {Manteiga}, {Massari},
  {Palaversa}, {Pancino}, {Riello}, {Ruz Mieres}, {Sanna}, {Santovena},
  {Sordo}, {Vallenari}, {Walton}, and {DPAC}]{2022arXiv220606215G}
{Gaia Collaboration}; {Montegriffo}, P.; {Bellazzini}, M.; {De Angeli}, F.;
  {Andrae}, R.; {Barstow}, M.A.; {Bossini}, D.; {Bragaglia}, A.; {Burgess},
  P.W.; {Cacciari}, C.;  et~al.
\newblock {Gaia Data Release 3: The Galaxy in your preferred colours. Synthetic
  photometry from Gaia low-resolution spectra}.
\newblock {\em arXiv} {\bf 2022},  {arXiv:2206.06215.} 


\bibitem[{Gaia Collaboration} et~al.(2016){Gaia Collaboration}, {Prusti}, {de
  Bruijne}, {Brown}, {Vallenari}, {Babusiaux}, {Bailer-Jones}, {Bastian},
  {Biermann}, {Evans}, {Eyer}, {Jansen}, {Jordi}, {Klioner}, {Lammers},
  {Lindegren}, {Luri}, {Mignard}, {Milligan}, {Panem}, {Poinsignon},
  {Pourbaix}, {Randich}, {Sarri}, {Sartoretti}, {Siddiqui}, {Soubiran},
  {Valette}, {van Leeuwen}, {Walton}, {Aerts}, {Arenou}, {Cropper}, {Drimmel},
  {H{\o}g}, {Katz}, {Lattanzi}, {O'Mullane}, {Grebel}, {Holland}, {Huc},
  {Passot}, {Bramante}, {Cacciari}, {Casta{\~n}eda}, {Chaoul}, {Cheek}, {De
  Angeli}, {Fabricius}, {Guerra}, {Hern{\'a}ndez}, {Jean-Antoine-Piccolo},
  {Masana}, {Messineo}, {Mowlavi}, {Nienartowicz}, {Ord{\'o}{\~n}ez-Blanco},
  {Panuzzo}, {Portell}, {Richards}, {Riello}, {Seabroke}, {Tanga},
  {Th{\'e}venin}, {Torra}, {Els}, {Gracia-Abril}, {Comoretto},
  {Garcia-Reinaldos}, {Lock}, {Mercier}, {Altmann}, {Andrae}, {Astraatmadja},
  {Bellas-Velidis}, {Benson}, {Berthier}, {Blomme}, {Busso}, {Carry},
  {Cellino}, {Clementini}, {Cowell}, {Creevey}, {Cuypers}, {Davidson}, {De
  Ridder}, {de Torres}, {Delchambre}, {Dell'Oro}, {Ducourant}, {Fr{\'e}mat},
  {Garc{\'\i}a-Torres}, {Gosset}, {Halbwachs}, {Hambly}, {Harrison}, {Hauser},
  {Hestroffer}, {Hodgkin}, {Huckle}, {Hutton}, {Jasniewicz}, {Jordan},
  {Kontizas}, {Korn}, {Lanzafame}, {Manteiga}, {Moitinho}, {Muinonen},
  {Osinde}, {Pancino}, {Pauwels}, {Petit}, {Recio-Blanco}, {Robin}, {Sarro},
  {Siopis}, {Smith}, {Smith}, {Sozzetti}, {Thuillot}, {van Reeven}, {Viala},
  {Abbas}, {Abreu Aramburu}, {Accart}, {Aguado}, {Allan}, {Allasia},
  {Altavilla}, {{\'A}lvarez}, {Alves}, {Anderson}, {Andrei}, {Anglada Varela},
  {Antiche}, {Antoja}, {Ant{\'o}n}, {Arcay}, {Atzei}, {Ayache}, {Bach},
  {Baker}, {Balaguer-N{\'u}{\~n}ez}, {Barache}, {Barata}, {Barbier}, {Barblan},
  {Baroni}, {Barrado y Navascu{\'e}s}, {Barros}, {Barstow}, {Becciani},
  {Bellazzini}, {Bellei}, {Bello Garc{\'\i}a}, {Belokurov}, {Bendjoya},
  {Berihuete}, {Bianchi}, {Bienaym{\'e}}, {Billebaud}, {Blagorodnova},
  {Blanco-Cuaresma}, {Boch}, {Bombrun}, {Borrachero}, {Bouquillon}, {Bourda},
  {Bouy}, {Bragaglia}, {Breddels}, {Brouillet}, {Br{\"u}semeister},
  {Bucciarelli}, {Budnik}, {Burgess}, {Burgon}, {Burlacu}, {Busonero}, {Buzzi},
  {Caffau}, {Cambras}, {Campbell}, {Cancelliere}, {Cantat-Gaudin}, {Carlucci},
  {Carrasco}, {Castellani}, {Charlot}, {Charnas}, {Charvet}, {Chassat},
  {Chiavassa}, {Clotet}, {Cocozza}, {Collins}, {Collins}, {Costigan}, {Crifo},
  {Cross}, {Crosta}, {Crowley}, {Dafonte}, {Damerdji}, {Dapergolas}, {David},
  {David}, {De Cat}, {de Felice}, {de Laverny}, {De Luise}, {De March}, {de
  Martino}, {de Souza}, {Debosscher}, {del Pozo}, {Delbo}, {Delgado},
  {Delgado}, {di Marco}, {Di Matteo}, {Diakite}, {Distefano}, {Dolding}, {Dos
  Anjos}, {Drazinos}, {Dur{\'a}n}, {Dzigan}, {Ecale}, {Edvardsson}, {Enke},
  {Erdmann}, {Escolar}, {Espina}, {Evans}, {Eynard Bontemps}, {Fabre},
  {Fabrizio}, {Faigler}, {Falc{\~a}o}, {Farr{\`a}s Casas}, {Faye}, {Federici},
  {Fedorets}, {Fern{\'a}ndez-Hern{\'a}ndez}, {Fernique}, {Fienga}, {Figueras},
  {Filippi}, {Findeisen}, {Fonti}, {Fouesneau}, {Fraile}, {Fraser}, {Fuchs},
  {Furnell}, {Gai}, {Galleti}, {Galluccio}, {Garabato}, {Garc{\'\i}a-Sedano},
  {Gar{\'e}}, {Garofalo}, {Garralda}, {Gavras}, {Gerssen}, {Geyer}, {Gilmore},
  {Girona}, {Giuffrida}, {Gomes}, {Gonz{\'a}lez-Marcos},
  {Gonz{\'a}lez-N{\'u}{\~n}ez}, {Gonz{\'a}lez-Vidal}, {Granvik}, {Guerrier},
  {Guillout}, {Guiraud}, {G{\'u}rpide}, {Guti{\'e}rrez-S{\'a}nchez}, {Guy},
  {Haigron}, {Hatzidimitriou}, {Haywood}, {Heiter}, {Helmi}, {Hobbs},
  {Hofmann}, {Holl}, {Holland }, {Hunt}, {Hypki}, {Icardi}, {Irwin}, {Jevardat
  de Fombelle}, {Jofr{\'e}}, {Jonker}, {Jorissen}, {Julbe}, {Karampelas},
  {Kochoska}, {Kohley}, {Kolenberg}, {Kontizas}, {Koposov}, {Kordopatis},
  {Koubsky}, {Kowalczyk}, {Krone-Martins}, {Kudryashova}, {Kull}, {Bachchan},
  {Lacoste-Seris}, {Lanza}, {Lavigne}, {Le Poncin-Lafitte}, {Lebreton},
  {Lebzelter}, {Leccia}, {Leclerc}, {Lecoeur-Taibi}, {Lemaitre}, {Lenhardt},
  {Leroux}, {Liao}, {Licata}, {Lindstr{\o}m}, {Lister}, {Livanou}, {Lobel},
  {L{\"o}ffler}, {L{\'o}pez}, {Lopez-Lozano}, {Lorenz}, {Loureiro},
  {MacDonald}, {Magalh{\~a}es Fernandes}, {Managau}, {Mann}, {Mantelet},
  {Marchal}, {Marchant}, {Marconi}, {Marie}, {Marinoni}, {Marrese},
  {Marschalk{\'o}}, {Marshall}, {Mart{\'\i}n-Fleitas}, {Martino}, {Mary},
  {Matijevi{\v{c}}}, {Mazeh}, {McMillan}, {Messina}, {Mestre}, {Michalik},
  {Millar}, {Miranda}, {Molina}, {Molinaro}, {Molinaro}, {Moln{\'a}r},
  {Moniez}, {Montegriffo}, {Monteiro}, {Mor}, {Mora}, {Morbidelli}, {Morel},
  {Morgenthaler}, {Morley}, {Morris}, {Mulone}, {Muraveva}, {Musella},
  {Narbonne}, {Nelemans}, {Nicastro}, {Noval}, {Ord{\'e}novic},
  {Ordieres-Mer{\'e}}, {Osborne}, {Pagani}, {Pagano}, {Pailler}, {Palacin},
  {Palaversa}, {Parsons}, {Paulsen}, {Pecoraro}, {Pedrosa}, {Pentik{\"a}inen},
  {Pereira}, {Pichon}, {Piersimoni}, {Pineau}, {Plachy}, {Plum}, {Poujoulet},
  {Pr{\v{s}}a}, {Pulone}, {Ragaini}, {Rago}, {Rambaux}, {Ramos-Lerate},
  {Ranalli}, {Rauw}, {Read}, {Regibo}, {Renk}, {Reyl{\'e}}, {Ribeiro},
  {Rimoldini}, {Ripepi}, {Riva}, {Rixon}, {Roelens}, {Romero-G{\'o}mez},
  {Rowell}, {Royer}, {Rudolph}, {Ruiz-Dern}, {Sadowski}, {Sagrist{\`a}
  Sell{\'e}s}, {Sahlmann}, {Salgado}, {Salguero}, {Sarasso}, {Savietto},
  {Schnorhk}, {Schultheis}, {Sciacca}, {Segol}, {Segovia}, {Segransan},
  {Serpell}, {Shih}, {Smareglia}, {Smart}, {Smith}, {Solano}, {Solitro},
  {Sordo}, {Soria Nieto}, {Souchay}, {Spagna}, {Spoto}, {Stampa}, {Steele},
  {Steidelm{\"u}ller}, {Stephenson}, {Stoev}, {Suess}, {S{\"u}veges}, {Surdej},
  {Szabados}, {Szegedi-Elek}, {Tapiador}, {Taris}, {Tauran}, {Taylor},
  {Teixeira}, {Terrett}, {Tingley}, {Trager}, {Turon}, {Ulla}, {Utrilla},
  {Valentini}, {van Elteren}, {Van Hemelryck}, {van Leeuwen}, {Varadi},
  {Vecchiato}, {Veljanoski}, {Via}, {Vicente}, {Vogt}, {Voss}, {Votruba},
  {Voutsinas}, {Walmsley}, {Weiler}, {Weingrill}, {Werner}, {Wevers},
  {Whitehead}, {Wyrzykowski}, {Yoldas}, {{\v{Z}}erjal}, {Zucker}, {Zurbach},
  {Zwitter}, {Alecu}, {Allen}, {Allende Prieto}, {Amorim},
  {Anglada-Escud{\'e}}, {Arsenijevic}, {Azaz}, {Balm}, {Beck}, {Bernstein},
  {Bigot}, {Bijaoui}, {Blasco}, {Bonfigli}, {Bono}, {Boudreault}, {Bressan},
  {Brown}, {Brunet}, {Bunclark}, {Buonanno}, {Butkevich}, {Carret}, {Carrion},
  {Chemin}, {Ch{\'e}reau}, {Corcione}, {Darmigny}, {de Boer}, {de Teodoro}, {de
  Zeeuw}, {Delle Luche}, {Domingues}, {Dubath}, {Fodor}, {Fr{\'e}zouls},
  {Fries}, {Fustes}, {Fyfe}, {Gallardo}, {Gallegos}, {Gardiol}, {Gebran},
  {Gomboc}, {G{\'o}mez}, {Grux}, {Gueguen}, {Heyrovsky}, {Hoar}, {Iannicola},
  {Isasi Parache}, {Janotto}, {Joliet}, {Jonckheere}, {Keil}, {Kim},
  {Klagyivik}, {Klar}, {Knude}, {Kochukhov}, {Kolka}, {Kos}, {Kutka}, {Lainey},
  {LeBouquin}, {Liu}, {Loreggia}, {Makarov}, {Marseille}, {Martayan},
  {Martinez-Rubi}, {Massart}, {Meynadier}, {Mignot}, {Munari}, {Nguyen},
  {Nordlander}, {Ocvirk}, {O'Flaherty}, {Olias Sanz}, {Ortiz}, {Osorio},
  {Oszkiewicz}, {Ouzounis}, {Palmer}, {Park}, {Pasquato}, {Peltzer}, {Peralta},
  {P{\'e}turaud}, {Pieniluoma}, {Pigozzi}, {Poels}, {Prat}, {Prod'homme},
  {Raison}, {Rebordao}, {Risquez}, {Rocca-Volmerange}, {Rosen}, {Ruiz-Fuertes},
  {Russo}, {Sembay}, {Serraller Vizcaino}, {Short}, {Siebert}, {Silva},
  {Sinachopoulos}, {Slezak}, {Soffel}, {Sosnowska}, {Strai{\v{z}}ys}, {ter
  Linden}, {Terrell}, {Theil}, {Tiede}, {Troisi}, {Tsalmantza}, {Tur},
  {Vaccari}, {Vachier}, {Valles}, {Van Hamme}, {Veltz}, {Virtanen}, {Wallut},
  {Wichmann}, {Wilkinson}, {Ziaeepour}, and {Zschocke}]{2016A&A...595A...1G}
{Gaia Collaboration}.; {Prusti}, T.; {de Bruijne}, J.H.J.; {Brown}, A.G.A.;
  {Vallenari}, A.; {Babusiaux}, C.; {Bailer-Jones}, C.A.L.; {Bastian}, U.;
  {Biermann}, M.; {Evans}, D.W.;  et~al.
\newblock {The Gaia mission}.  \emph{Astron. Astrophys.}
\newblock  {\bf 2016}, {\em 595},~ {A1.} 
\newblock {\url{https://doi.org/10.1051/0004-6361/201629272}}.

\bibitem[{Skrutskie} et~al.(2006){Skrutskie}, {Cutri}, {Stiening}, {Weinberg},
  {Schneider}, {Carpenter}, {Beichman}, {Capps}, {Chester}, {Elias}, {Huchra},
  {Liebert}, {Lonsdale}, {Monet}, {Price}, {Seitzer}, {Jarrett}, {Kirkpatrick},
  {Gizis}, {Howard}, {Evans}, {Fowler}, {Fullmer}, {Hurt}, {Light}, {Kopan},
  {Marsh}, {McCallon}, {Tam}, {Van Dyk}, and {Wheelock}]{2006AJ....131.1163S}
{Skrutskie}, M.F.; {Cutri}, R.M.; {Stiening}, R.; {Weinberg}, M.D.;
  {Schneider}, S.; {Carpenter}, J.M.; {Beichman}, C.; {Capps}, R.; {Chester},
  T.; {Elias}, J.;  et~al.
\newblock {The Two Micron All Sky Survey (2MASS)}. \emph{Astron. J.}
\newblock  {\bf 2006}, {\em 131},~1163--1183.
\newblock {\url{https://doi.org/10.1086/498708}}.

\bibitem[{Wright} et~al.(2010){Wright}, {Eisenhardt}, {Mainzer}, {Ressler},
  {Cutri}, {Jarrett}, {Kirkpatrick}, {Padgett}, {McMillan}, {Skrutskie},
  {Stanford}, {Cohen}, {Walker}, {Mather}, {Leisawitz}, {Gautier}, {McLean},
  {Benford}, {Lonsdale}, {Blain}, {Mendez}, {Irace}, {Duval}, {Liu}, {Royer},
  {Heinrichsen}, {Howard}, {Shannon}, {Kendall}, {Walsh}, {Larsen}, {Cardon},
  {Schick}, {Schwalm}, {Abid}, {Fabinsky}, {Naes}, and
  {Tsai}]{2010AJ....140.1868W}
{Wright}, E.L.; {Eisenhardt}, P.R.M.; {Mainzer}, A.K.; {Ressler}, M.E.;
  {Cutri}, R.M.; {Jarrett}, T.; {Kirkpatrick}, J.D.; {Padgett}, D.; {McMillan},
  R.S.; {Skrutskie}, M.;  et~al.
\newblock {The Wide-field Infrared Survey Explorer (WISE): Mission Description
  and Initial On-orbit Performance}. \emph{Astron. J.}
\newblock  {\bf 2010}, {\em 140},~1868-- {1881.} 
\newblock {\url{https://doi.org/10.1088/0004-6256/140/6/1868}}.

\bibitem[{Fouqu{\'e}} et~al.(2000){Fouqu{\'e}}, {Chevallier}, {Cohen},
  {Galliano}, {Loup}, {Alard}, {de Batz}, {Bertin}, {Borsenberger}, {Cioni},
  {Copet}, {Dennefeld}, {Derriere}, {Deul}, {Duc}, {Egret}, {Epchtein},
  {Forveille}, {Garz{\'o}n}, {Habing}, {Hron}, {Kimeswenger}, {Lacombe}, {Le
  Bertre}, {Mamon}, {Omont}, {Paturel}, {Pau}, {Persi}, {Robin}, {Rouan},
  {Schultheis}, {Simon}, {Tiph{\`e}ne}, {Vauglin}, and
  {Wagner}]{2000A&AS..141..313F}
{Fouqu{\'e}}, P.; {Chevallier}, L.; {Cohen}, M.; {Galliano}, E.; {Loup}, C.;
  {Alard}, C.; {de Batz}, B.; {Bertin}, E.; {Borsenberger}, J.; {Cioni}, M.R.;
  et~al.
\newblock {An absolute calibration of DENIS (deep near infrared southern sky
  survey)}. \emph{Astron. Astrophys. Suppl. Ser.}
\newblock  {\bf 2000}, {\em 141},~313--317.
\newblock {\url{https://doi.org/10.1051/aas:2000123}}.

\bibitem[{Epchtein} et~al.(1999){Epchtein}, {Deul}, {Derriere}, {Borsenberger},
  {Egret}, {Simon}, {Alard}, {Bal{\'a}zs}, {de Batz}, {Cioni}, {Copet},
  {Dennefeld}, {Forveille}, {Fouqu{\'e}}, {Garz{\'o}n}, {Habing}, {Holl},
  {Hron}, {Kimeswenger}, {Lacombe}, {Le Bertre}, {Loup}, {Mamon}, {Omont},
  {Paturel}, {Persi}, {Robin}, {Rouan}, {Tiph{\`e}ne}, {Vauglin}, and
  {Wagner}]{1999A&A...349..236E}
{Epchtein}, N.; {Deul}, E.; {Derriere}, S.; {Borsenberger}, J.; {Egret}, D.;
  {Simon}, G.; {Alard}, C.; {Bal{\'a}zs}, L.G.; {de Batz}, B.; {Cioni}, M.R.;
  et~al.
\newblock {A preliminary database of DENIS point sources}. \emph{Astron. Astrophys.}
\newblock {\bf 1999}, {\em 349},~236--242.

\bibitem[{Mainzer} et~al.(2014){Mainzer}, {Bauer}, {Cutri}, {Grav}, {Masiero},
  {Beck}, {Clarkson}, {Conrow}, {Dailey}, {Eisenhardt}, {Fabinsky},
  {Fajardo-Acosta}, {Fowler}, {Gelino}, {Grillmair}, {Heinrichsen}, {Kendall},
  {Kirkpatrick}, {Liu}, {Masci}, {McCallon}, {Nugent}, {Papin}, {Rice},
  {Royer}, {Ryan}, {Sevilla}, {Sonnett}, {Stevenson}, {Thompson}, {Wheelock},
  {Wiemer}, {Wittman}, {Wright}, and {Yan}]{2014ApJ...792...30M}
{Mainzer}, A.; {Bauer}, J.; {Cutri}, R.M.; {Grav}, T.; {Masiero}, J.; {Beck},
  R.; {Clarkson}, P.; {Conrow}, T.; {Dailey}, J.; {Eisenhardt}, P.;  et~al.
\newblock {Initial Performance of the NEOWISE Reactivation Mission}. \emph{Astrophys. J.}
\newblock {\bf 2014}, {\em 792},~ {30.} 
\newblock {\url{https://doi.org/10.1088/0004-637X/792/1/30}}.

\bibitem[Pakhomov et~al.(2017)Pakhomov, Piskunov, and
  Ryabchikova]{pakhomov2017vald3currentdevelopments}
Pakhomov, Y.; Piskunov, N.; Ryabchikova, T.
\newblock VALD3: Current developments.  \emph{arXiv} \textbf{2017},  {arXiv:1710.10854.} 

 

\bibitem[Wehrhahn et~al.(2023)Wehrhahn, Piskunov, and
  Ryabchikova]{Wehrhahn_2023}
Wehrhahn, A.; Piskunov, N.; Ryabchikova, T.
\newblock PySME: Spectroscopy Made Easier. \emph{Astron. Astrophys.}
\newblock {\bf 2023}, {\em 671},~A171.
\newblock {\url{https://doi.org/10.1051/0004-6361/202244482}}.

\bibitem[Castelli and Kurucz(2004)]{castelli2004newgridsatlas9model}
Castelli, F.; Kurucz, R.L.
\newblock New Grids of ATLAS9 Model Atmospheres. \emph{arXiv}  \textbf{2004},  {arXiv:astro-ph/0405087.} 

\bibitem[Asplund et~al.(2021)Asplund, Amarsi, and Grevesse]{Asplund_2021}
Asplund, M.; Amarsi, A.M.; Grevesse, N.
\newblock The chemical make-up of the Sun: A 2020 vision. \emph{Astron. Astrophys.}
\newblock {\bf 2021}, {\em 653},~A141.
\newblock {\url{https://doi.org/10.1051/0004-6361/202140445}}.

\bibitem[Kochukhov(2018)]{kochukhov2018binmag}
Kochukhov, O.
\newblock \emph{BinMag: Widget for Comparing Stellar Observed with Theoretical
  Spectra};
\newblock { {Astrophysics Source Code Library:} 
}, {2018}; \linebreak  p. ascl--1805.

\bibitem[Tkachenko(2015)]{Tkachenko_2015}
Tkachenko, A.
\newblock Grid search in stellar parameters: A software for spectrum analysis
  of single stars and binary systems. \emph{Astron. Astrophys.}
\newblock {\bf 2015}, {\em 581},~A129.
\newblock {\url{https://doi.org/10.1051/0004-6361/201526513}}.

\bibitem[Dileep et~al.(2025)Dileep, Joshi, Alexeeva, Kochukhov, Semenko,
  De~Cat, Zúñiga-Fernández, Trust, Pollard, Crause, Barkaoui, Gillon, Jehin,
  and Rathore]{10.1093/mnras/staf1247}
Dileep, A.; Joshi, S.; Alexeeva, S.; Kochukhov, O.; Semenko, E.; De~Cat, P.;
  Zúñiga-Fernández, S.; Trust, O.; Pollard, K.; Crause, L.;  et~al.
\newblock Chemical abundances and doppler imaging of the Ap Si/He-wk star
  HD 100357.
\newblock {\em Mon. Not. R. Astron. Soc.} {\bf 2025},
   {staf1247.} 
\newblock {\url{https://doi.org/10.1093/mnras/staf1247}}.

\bibitem[Mashonkina et~al.(2011)Mashonkina, Gehren, Shi, Korn, and
  Grupp]{Mashonkina2011}
Mashonkina, L.; Gehren, T.; Shi, J.R.; Korn, A.J.; Grupp, F.
\newblock {Non-LTE line formation for Fe and Ca in late-type stars--I.
  Statistical equilibrium of neutral and singly-ionised species}. \emph{Astron. Astrophys.}
\newblock  {\bf 2011}, {\em 528},~A87.

\bibitem[Przybilla et~al.(2001)Przybilla, Butler, Becker, Kudritzki, and
  Venn]{Przybilla2001}
Przybilla, N.; Butler, K.; Becker, S.R.; Kudritzki, R.P.; Venn, K.A.
\newblock {Non-LTE line-formation for hydrogen and helium in early-type stars---II. The line spectrum of the B8 IV star HD 160762}. \emph{Astron. Astrophys.}
\newblock {\bf 2001}, {\em 369},~1009.

\bibitem[Ryabchikova et~al.(2005)Ryabchikova, Leone, and
  Kochukhov]{Ryabchikova2005}
Ryabchikova, T.; Leone, F.; Kochukhov, O.
\newblock {Vertical abundance stratification in peculiar A-type stars: Silicon
  in 10 Aql}. \emph{Astron. Astrophys.}
\newblock {\bf 2005}, {\em 438},~973.

\bibitem[Romanovskaya et~al.(2024)Romanovskaya, Kochukhov, and
  Ryabchikova]{Romanovskaya2024}
Romanovskaya, E.V.; Kochukhov, O.; Ryabchikova, T.
\newblock {Stratification and Non-LTE Effects in the Slowly Rotating Ap Star BD
  +00°1659}. \emph{Astron. Astrophys.}
\newblock {\bf 2024}, {\em 670},~A45.

\bibitem[Kochukhov et~al.(2007)Kochukhov, Ryabchikova, Weiss, and
  Piskunov]{Kochukhov2007}
Kochukhov, O.; Ryabchikova, T.; Weiss, W.W.; Piskunov, N.
\newblock {Discovery of a Magnetic Field in the Rapidly Oscillating Ap Star HD
  99563}. \emph{Nature}
\newblock {\bf 2007}, {\em 450},~633.

\bibitem[Hubeny and Lanz(1995)]{Hubeny1995}
Hubeny, I.; Lanz, T.
\newblock {Non-LTE line-blanketed model atmospheres of hot stars. 1: Hybrid
  complete linearization/accelerated lambda iteration method}. \emph{Astrophys. J.}
\newblock {\bf 1995}, {\em 439},~875.

\bibitem[Butler and Giddings(1985)]{Butler1985}
Butler, K.; Giddings, J.
\newblock {\emph{Newsletter on Analysis of Astronomical Spectra, No. 9}};
\newblock Technical Report; University of London: London, UK, 1985.

\bibitem[{Bressan} et~al.(2012){Bressan}, {Marigo}, {Girardi}, {Salasnich},
  {Dal Cero}, {Rubele}, and {Nanni}]{2012MNRAS.427..127B}
{Bressan}, A.; {Marigo}, P.; {Girardi}, L.; {Salasnich}, B.; {Dal Cero}, C.;
  {Rubele}, S.; {Nanni}, A.
\newblock {PARSEC: Stellar tracks and isochrones with the PAdova and TRieste
  Stellar Evolution Code}. \emph{Mon. Not. R. Astron. Soc.}
\newblock {\bf 2012}, {\em 427},~127-- {145.} 
\newblock {\url{https://doi.org/10.1111/j.1365-2966.2012.21948.x}}.

\bibitem[{Hubrig} et~al.(2000){Hubrig}, {North}, and
  {Mathys}]{2000ApJ...539..352H}
{Hubrig}, S.; {North}, P.; {Mathys}, G.
\newblock {Magnetic AP Stars in the Hertzsprung-Russell Diagram}. \emph{Astrophys. J.}
\newblock {\bf 2000}, {\em 539},~352--363. 
\newblock {\url{https://doi.org/10.1086/309189}}.

\bibitem[{Torres}(2010)]{2010torres}
{Torres}, G.
\newblock {On the Use of Empirical Bolometric Corrections for Stars}. \emph{Astron. J.}
\newblock {\bf 2010}, {\em 140},~1158-- {1162.} 
\newblock {\url{https://doi.org/10.1088/0004-6256/140/5/1158}}.

\bibitem[{Gaia Collaboration} et~al.(2021){Gaia Collaboration}, {Brown},
  {Vallenari}, {Prusti}, {de Bruijne}, {Babusiaux}, {Biermann}, {Creevey},
  {Evans}, {Eyer}, {Hutton}, {Jansen}, {Jordi}, {Klioner}, {Lammers},
  {Lindegren}, {Luri}, {Mignard}, {Panem}, {Pourbaix}, {Randich}, {Sartoretti},
  {Soubiran}, {Walton}, {Arenou}, {Bailer-Jones}, {Bastian}, {Cropper},
  {Drimmel}, {Katz}, {Lattanzi}, {van Leeuwen}, {Bakker}, {Cacciari},
  {Casta{\~n}eda}, {De Angeli}, {Ducourant}, {Fabricius}, {Fouesneau},
  {Fr{\'e}mat}, {Guerra}, {Guerrier}, {Guiraud}, {Jean-Antoine Piccolo},
  {Masana}, {Messineo}, {Mowlavi}, {Nicolas}, {Nienartowicz}, {Pailler},
  {Panuzzo}, {Riclet}, {Roux}, {Seabroke}, {Sordo}, {Tanga}, {Th{\'e}venin},
  {Gracia-Abril}, {Portell}, {Teyssier}, {Altmann}, {Andrae}, {Bellas-Velidis},
  {Benson}, {Berthier}, {Blomme}, {Brugaletta}, {Burgess}, {Busso}, {Carry},
  {Cellino}, {Cheek}, {Clementini}, {Damerdji}, {Davidson}, {Delchambre},
  {Dell'Oro}, {Fern{\'a}ndez-Hern{\'a}ndez}, {Galluccio}, {Garc{\'\i}a-Lario},
  {Garcia-Reinaldos}, {Gonz{\'a}lez-N{\'u}{\~n}ez}, {Gosset}, {Haigron},
  {Halbwachs}, {Hambly}, {Harrison}, {Hatzidimitriou}, {Heiter},
  {Hern{\'a}ndez}, {Hestroffer}, {Hodgkin}, {Holl}, {Jan{\ss}en}, {Jevardat de
  Fombelle}, {Jordan}, {Krone-Martins}, {Lanzafame}, {L{\"o}ffler}, {Lorca},
  {Manteiga}, {Marchal}, {Marrese}, {Moitinho}, {Mora}, {Muinonen}, {Osborne},
  {Pancino}, {Pauwels}, {Petit}, {Recio-Blanco}, {Richards}, {Riello},
  {Rimoldini}, {Robin}, {Roegiers}, {Rybizki}, {Sarro}, {Siopis}, {Smith},
  {Sozzetti}, {Ulla}, {Utrilla}, {van Leeuwen}, {van Reeven}, {Abbas}, {Abreu
  Aramburu}, {Accart}, {Aerts}, {Aguado}, {Ajaj}, {Altavilla}, {{\'A}lvarez},
  {{\'A}lvarez Cid-Fuentes}, {Alves}, {Anderson}, {Anglada Varela}, {Antoja},
  {Audard}, {Baines}, {Baker}, {Balaguer-N{\'u}{\~n}ez}, {Balbinot}, {Balog},
  {Barache}, {Barbato}, {Barros}, {Barstow}, {Bartolom{\'e}}, {Bassilana},
  {Bauchet}, {Baudesson-Stella}, {Becciani}, {Bellazzini}, {Bernet}, {Bertone},
  {Bianchi}, {Blanco-Cuaresma}, {Boch}, {Bombrun}, {Bossini}, {Bouquillon},
  {Bragaglia}, {Bramante}, {Breedt}, {Bressan}, {Brouillet}, {Bucciarelli},
  {Burlacu}, {Busonero}, {Butkevich}, {Buzzi}, {Caffau}, {Cancelliere},
  {C{\'a}novas}, {Cantat-Gaudin}, {Carballo}, {Carlucci}, {Carnerero},
  {Carrasco}, {Casamiquela}, {Castellani}, {Castro-Ginard}, {Castro Sampol},
  {Chaoul}, {Charlot}, {Chemin}, {Chiavassa}, {Cioni}, {Comoretto}, {Cooper},
  {Cornez}, {Cowell}, {Crifo}, {Crosta}, {Crowley}, {Dafonte}, {Dapergolas},
  {David}, and {David}]{2021A&A...649A...1G}
{Gaia Collaboration}.; {Brown}, A.G.A.; {Vallenari}, A.; {Prusti}, T.; {de
  Bruijne}, J.H.J.; {Babusiaux}, C.; {Biermann}, M.; {Creevey}, O.L.; {Evans},
  D.W.; {Eyer}, L.;  et~al.
\newblock {Gaia Early Data Release 3. Summary of the contents and survey
  properties}. \emph{Astron. Astrophys.}
\newblock {\bf 2021}, {\em 649},~ {A1.} 
\newblock {\url{https://doi.org/10.1051/0004-6361/202039657}}.

\bibitem[{McEvoy} et~al.(2015){McEvoy}, {Smoker}, {Dufton}, {Smith}, {Kennedy},
  {Keenan}, {Lambert}, {Welty}, and {Lauroesch}]{2015MNRAS.451.1396M}
{McEvoy}, C.M.; {Smoker}, J.V.; {Dufton}, P.L.; {Smith}, K.T.; {Kennedy}, M.B.;
  {Keenan}, F.P.; {Lambert}, D.L.; {Welty}, D.E.; {Lauroesch}, J.T.
\newblock {Early-type stars observed in the ESO UVES Paranal Observatory
  Project - V. Time-variable interstellar absorption}. \emph{Mon. Not. R. Astron. Soc.}
\newblock  {\bf 2015}, {\em 451},~1396-- {1412.} 
\newblock {\url{https://doi.org/10.1093/mnras/stv945}}.

\bibitem[{Glagolevskij}(2019)]{2019AstBu..74...66G}
{Glagolevskij}, Y.V.
\newblock {On Properties of Main Sequence Magnetic Stars}.
\newblock {\em Astrophys. Bull.} {\bf 2019}, {\em 74},~66--79.
\newblock {\url{https://doi.org/10.1134/S1990341319010073}}.

\bibitem[{Borisov} et~al.(2023){Borisov}, {Chilingarian}, {Rubtsov}, {Ledoux},
  {Melo}, {Grishin}, {Katkov}, {Goradzhanov}, {Afanasiev}, {Kasparova}, and
  {Saburova}]{2023ApJS..266...11B}
{Borisov}, S.B.; {Chilingarian}, I.V.; {Rubtsov}, E.V.; {Ledoux}, C.; {Melo},
  C.; {Grishin}, K.A.; {Katkov}, I.Y.; {Goradzhanov}, V.S.; {Afanasiev}, A.V.;
  {Kasparova}, A.V.;  et~al.
\newblock {New Generation Stellar Spectral Libraries in the Optical and
  Near-infrared. I. The Recalibrated UVES-POP Library for Stellar Population
  Synthesis}.  \emph{Astrophys. J. Suppl. Ser.}
\newblock  {\bf 2023}, {\em 266},~ {11.} 
\newblock {\url{https://doi.org/10.3847/1538-4365/acc321}}.

\bibitem[{van Leeuwen}(2007)]{2007A&A...474..653V}
{van Leeuwen}, F.
\newblock {Validation of the new Hipparcos reduction}. \emph{Astron. Astrophys.}
\newblock {\bf 2007}, {\em 474},~653-- {664.} 
\newblock {\url{https://doi.org/10.1051/0004-6361:20078357}}.

\bibitem[{Bicz} et~al.(2022){Bicz}, {Falewicz}, {Pietras}, {Siarkowski}, and
  {Pre{\'s}}]{Bicz2022ApJ...935..102B}
{Bicz}, K.; {Falewicz}, R.; {Pietras}, M.; {Siarkowski}, M.; {Pre{\'s}}, P.
\newblock {Starspot Modeling and Flare Analysis on Selected Main-sequence
  M-type Stars}. \emph{Astrophys. J.}
\newblock  {\bf 2022}, {\em 935},~ {102.} 
\newblock {\url{https://doi.org/10.3847/1538-4357/ac7ab3}}.

\bibitem[{Notsu} et~al.(2013){Notsu}, {Shibayama}, {Maehara}, {Notsu}, {Nagao},
  {Honda}, {Ishii}, {Nogami}, and {Shibata}]{2013ApJ...771..127N}
{Notsu}, Y.; {Shibayama}, T.; {Maehara}, H.; {Notsu}, S.; {Nagao}, T.; {Honda},
  S.; {Ishii}, T.T.; {Nogami}, D.; {Shibata}, K.
\newblock {Superflares on Solar-type Stars Observed with Kepler II. Photometric
  Variability of Superflare-generating Stars: A Signature of Stellar Rotation
  and Starspots}. \emph{Astrophys. J.}
\newblock  {\bf 2013}, {\em 771},~ {127.} 
\newblock {\url{https://doi.org/10.1088/0004-637X/771/2/127}}.

\bibitem[{Shibata} et~al.(2013){Shibata}, {Isobe}, {Hillier}, {Choudhuri},
  {Maehara}, {Ishii}, {Shibayama}, {Notsu}, {Notsu}, {Nagao}, {Honda}, and
  {Nogami}]{2013PASJ...65...49S}
{Shibata}, K.; {Isobe}, H.; {Hillier}, A.; {Choudhuri}, A.R.; {Maehara}, H.;
  {Ishii}, T.T.; {Shibayama}, T.; {Notsu}, S.; {Notsu}, Y.; {Nagao}, T.;
  et~al.
\newblock {Can Superflares Occur on Our Sun?} \emph{Publ. Astron. Soc. Jpn.}
\newblock  {\bf 2013}, {\em 65},~ {49.} 
\newblock {\url{https://doi.org/10.1093/pasj/65.3.49}}.

\bibitem[{Notsu} et~al.(2019){Notsu}, {Maehara}, {Honda}, {Hawley},
  {Davenport}, {Namekata}, {Notsu}, {Ikuta}, {Nogami}, and
  {Shibata}]{2019ApJ...876...58N}
{Notsu}, Y.; {Maehara}, H.; {Honda}, S.; {Hawley}, S.L.; {Davenport}, J.R.A.;
  {Namekata}, K.; {Notsu}, S.; {Ikuta}, K.; {Nogami}, D.; {Shibata}, K.
\newblock {Do Kepler Superflare Stars Really Include Slowly Rotating Sun-like
  Stars?{\textemdash}Results Using APO 3.5 m Telescope Spectroscopic
  Observations and Gaia-DR2 Data}. \emph{Astrophys. J.}
\newblock  {\bf 2019}, {\em 876},~ {58.} 
\newblock {\url{https://doi.org/10.3847/1538-4357/ab14e6}}.

\bibitem[{Walkowicz} et~al.(2013){Walkowicz}, {Basri}, and
  {Valenti}]{2013ApJS..205...17W}
{Walkowicz}, L.M.; {Basri}, G.; {Valenti}, J.A.
\newblock {The Information Content in Analytic Spot Models of Broadband
  Precision Light Curves}. \emph{Astrophys. J. Suppl. Ser.}
\newblock  {\bf 2013}, {\em 205},~ {17.} 
\newblock {\url{https://doi.org/10.1088/0067-0049/205/2/17}}.

\bibitem[{Leone} and {Catalano}(1991)]{1991A&A...242..199L}
{Leone}, F.; {Catalano}, F.A.
\newblock {The overall flux distribution of magnetic chemically peculiar
  stars.} \emph{Astron. Astrophys.}
\newblock {\bf 1991}, {\em 242},~199.

\bibitem[Bailey et~al.(2014)Bailey, Landstreet, and Bagnulo]{Bailey_2014}
Bailey, J.D.; Landstreet, J.D.; Bagnulo, S.
\newblock Discovery of secular variations in the atmospheric abundances of
  magnetic Ap stars. \emph{Astron. Astrophys.}
\newblock {\bf 2014}, {\em 561},~A147.
\newblock {\url{https://doi.org/10.1051/0004-6361/201322853}}.

\bibitem[Przybilla(2005)]{Przybilla2005}
Przybilla, N.
\newblock Non-LTE line formation for neutral helium in early-type stars. \emph{Astron. Astrophys.}
\newblock {\bf 2005}, {\em 443},~293--303.
\newblock {\url{https://doi.org/10.1051/0004-6361:20053570}}.

\bibitem[Nieva and Przybilla(2012)]{NievaPrzybilla2012}
Nieva, M.F.; Przybilla, N.
\newblock Present-day cosmic abundances: A comprehensive study of nearby early
  B-type stars. \emph{Astron. Astrophys.}
\newblock {\bf 2012}, {\em 539},~A143.
\newblock {\url{https://doi.org/10.1051/0004-6361/201118158}}.

\bibitem[Lind et~al.(2012)Lind, Bergemann, and Asplund]{Lind2012}
Lind, K.; Bergemann, M.; Asplund, M.
\newblock Non-LTE line formation of Fe in late-type stars---{I}. Standard stars
  with 1D and 3D model atmospheres.  \emph{ {Mon. Not. R. Astron. Soc.} 
 }
\newblock {\bf 2012}, {\em 427},~50-- {60.} 
\newblock {\url{https://doi.org/10.1111/j.1365-2966.2012.21687.x}}.

\bibitem[Mashonkina et~al.(2016)Mashonkina, Sitnova, and
  Pakhomov]{Mashonkina2016}
Mashonkina, L.I.; Sitnova, T.N.; Pakhomov, Y.V.
\newblock Non-LTE abundance determinations for silicon in A--B-type stellar
  atmospheres.
\newblock {\em Astron. Lett.} {\bf 2016}, {\em 42},~606--617.
\newblock {\url{https://doi.org/10.1134/S1063773716080034}}.

\bibitem[Sitnova et~al.(2020)Sitnova, Mashonkina, and Pakhomov]{Sitnova2020}
Sitnova, T.N.; Mashonkina, L.I.; Pakhomov, Y.V.
\newblock Non-LTE line formation for Si{i--iii} in A--B stars and the
  origin of Siii emission in \textit{iota}\,Her. \emph{ {Mon. Not. R. Astron. Soc.} 
  }
\newblock {\bf 2020}, {\em 493},~6095--6108.
\newblock {\url{https://doi.org/10.1093/mnras/staa598}}.

\bibitem[Mashonkina et~al.(2005)Mashonkina, Ryabchikova, and
  Ryabtsev]{Mashonkina2005}
Mashonkina, L.; Ryabchikova, T.; Ryabtsev, A.
\newblock NLTE ionisation equilibrium of {Nd\,\textsc{ii}} and
  {Nd\,\textsc{iii}} in cool A and Ap stars. \emph{Astron. Astrophys.}
\newblock {\bf 2005}, {\em 441},~309--318.
\newblock {\url{https://doi.org/10.1051/0004-6361:20053085}}.

\bibitem[Ryabchikova et~al.(2015)Ryabchikova, Piskunov, and
  Kurucz]{Ryabchikova2015}
Ryabchikova, T.; Piskunov, N.; Kurucz, R.L.
\newblock A database of rare-earth element line data and its application to
  Ap-star spectra. \emph{ {Mon. Not. R. Astron. Soc.} 
  }
\newblock {\bf 2015}, {\em 447},~2046--2055.
\newblock {\url{https://doi.org/10.1093/mnras/stu2575}}.

\bibitem[{Bailey, J. D.} and {Landstreet, J. D.}(2013)]{refId0}
{Bailey, J.D.}; {Landstreet, J.D.}
\newblock Abundances determined using Si\,{\sc ii} and Si\,{\sc iii} in B-type
  stars: evidence for stratification. \emph{Astron. Astrophys.}
\newblock {\bf 2013}, {\em 551},~A30.
\newblock {\url{https://doi.org/10.1051/0004-6361/201220671}}.

\end{thebibliography}

\PublishersNote{}
\end{adjustwidth}
\end{document}